\documentclass{article}

\usepackage{arxiv}

\usepackage[utf8]{inputenc} 
\usepackage[T1]{fontenc}    
\usepackage{hyperref}       
\usepackage{url}            
\usepackage{booktabs}       
\usepackage{amsfonts}       
\usepackage{nicefrac}       
\usepackage{microtype}      
\usepackage{graphicx}
\usepackage{natbib}
\usepackage{doi}

\usepackage{amssymb}
\usepackage{latexsym}
\usepackage{epsfig}
\usepackage{amsmath}

\def\tends{\rightarrow}

\def\ZZ{\mathbb Z}

\def\EE{\mathbb E}

\def\xx{\mathbf{X}}
\def\yy{\mathbf{Y}}

\def\uu{\mathbf{U}}

\newcommand{\pkg}[1]{\textbf{#1}}
\newcommand{\proglang}[1]{\textsc{#1}}
\newcommand{\code}[1]{\texttt{#1}}

\title{\pkg{Ecce Signum}: An \proglang{R} Package for Multivariate Signal Extraction and Time Series Analysis}

\date{} 					

\author{ Tucker McElroy\\U.S. Census Bureau \\
  4600 Silver Hill Road, Washington, DC 20233\\
\texttt{tucker.s.mcelroy@census.gov} \And 
        James Livsey\\U.S. Census Bureau \\
         4600 Silver Hill Road, Washington, DC 20233\\
        \texttt{james.a.livsey@census.gov} 
}

\renewcommand{\shorttitle}{\pkg{Ecce Signum}}

\hypersetup{
pdftitle={\pkg{Ecce Signum}: An \proglang{R} Package for Multivariate Signal Extraction and Time Series Analysis},
pdfsubject={},
pdfauthor={Tucker S.~McElroy, James A.~Livsey},
pdfkeywords={Entropy maximization, Filtering, Missing values, Ragged edge, Seasonality, Signal extraction},
}

\begin{document}
\maketitle

\begin{abstract}
 The package provides multivariate time series models  for structural analysis, allowing one to extract
 latent signals such as trends or seasonality.  Models are fitted using maximum likelihood estimation, allowing for
 non-stationarity, fixed regression effects, and ragged-edge missing values.   Simple types of extreme values can be
 corrected using the device of entropy maximization.   Model adequacy is assessed through residual diagnostics,
 and model-based signal extraction filters can be assessed in time domain and frequency domain.  
Extracted signals are produced with uncertainty measures that account for sample edge effects and missing values, and
  the signals (as well as the original time series) can be forecasted. 
\end{abstract}


\section{Introduction}

The \proglang{R} package \pkg{Ecce Signum}\footnote{Latin for ``Behold the Proof"; 
cf. Falstaff's speech in Henry IV, Part 1 (Act 2, Scene 4).},
  or \pkg{sigex} for short, developed organically from \proglang{R} code for multivariate
 signal extraction required for the empirical applications of 
 \cite{mcelroy2015signal}.
   This methodology employed direct matrix formulas
 for signal extraction, which has certain advantages over the use of a state space smoother when the sample size is small.  
  For further applications it became necessary to extend the models to handle
   diverse types of co-integration effects, such as common trends and common seasonality,
 and the resulting methods were summarized in \cite{mcelroy2017multivariate}.  However, at the same time it was recognized that extending the
 non-state space methodology to longer samples would be advantageous, given that general forms of integrated processes can be
 handled, including non-Markovian processes and long memory processes not amenable to a state space embedding.  
  Algorithmic work   to make feasible these extensions is discussed in \cite{mcelroy2018recursive}  and  \cite{McElroyCasting}; 
   extensions to  model estimation are discussed in 
   \cite{mcelroy2018inverse}.
  
The result of these developments is a flexible set of tools for the analysis of multivariate time series.  
 Time series data with some or all
 of the following attributes can be analyzed with \pkg{Ecce Signum}:
\begin{itemize}
\item Multivariate  time series, or univariate time series with embedded multivariate structure
\item Fixed effects, such as holiday patterns
\item Aberrancies, such as additive outliers
\item Complex dynamics, including trend, seasonality, and business cycles
\item Seasonal cycles of diverse period (e.g., weekly and annual effects)
\item Linked dynamics, such as common trends and common seasonality
\item Ragged edge missing values
\item Changes in sampling frequency 
\end{itemize}
 Highly exotic time series cannot be handled appropriately with \pkg{Ecce Signum}.    Data that can be easily linearized 
 (i.e., rendered as a linear time series, possibly with Gaussian marginals) through
 transformations, regression, and differencing is appropriate for analysis.  
We have in mind several possible objectives of analysis, which the methodological tools can address:
\begin{itemize}
\item Modeling of the data via reduction to a maximum entropy residual
\item Identification and estimation of fixed effects
\item Linearization of the data via maximum entropy transformations
\item Extraction of underlying signals, which may be common, such as trends, seasonals, and cycles
\item Imputation of missing values
\item Forecasting and aftcasting 
\item Quantification of uncertainty in extractions
\end{itemize}
  The modeling philosophy is informed by the concept of entropy, and is discussed in \cite{McElroyPenny2019}.
 Tools are available to compare fitted models and evaluate their performance, 
 and two methods of fitting are presented: maximization of
 the Gaussian likelihood utilizing a proper factorization, and method-of-moments 
 --  which  is faster and more appropriate for high-dimensional  parameter spaces.   

Modeling of univariate structural time series is well covered in the \proglang{R} package world. 
In fact, base \proglang{R} includes the function \code{StructTS}, which fits a univariate structural time series model via maximizing the Gaussian likelihood. 
If an analyst is willing to put their model formulation into the state-space framework and make use of the Kalman filter and smooth functions, there exist many packages that can estimate such a specification, viz.,
\pkg{dlm} \citep{petris2010Rpackage}, \pkg{dse} (focuses on ARMA models) \citep{gilbert2020Rpackage}, \pkg{bsts} (Bayesian implementation) \citep{scott2020bsts}, and \pkg{stsm} \citep{lopez2016stsm}. 
There do exist some packages for processing multivariate structural time series.  
The \pkg{MARSS} package \citep{holmes2012marss, holmes2020Rpackage} adopts a state space approach to fitting and filtering algorithms. Additionally, \pkg{KFAS}  (Kalman Filter and Smoother for Exponential Family State Space Models, \cite{helske2016kfas}) can estimate multivariate time series models. 
\pkg{MARSS} uses \pkg{KFAS}, and implements more stable filtering/smoothing algorithms. 
\pkg{KFAS} on the other hand, allows users to consider non-Gaussian time series. 
The \pkg{MARSS} package currently can estimate a multivariate state-space model specification with seasonal components using either an EM algorithm or 
 nonlinear optimization via BFGS. 
 (However, it is noted in the user guide that the EM algorithm can be challenging for multivariate models.)
 
\pkg{Ecce Signum} does not rely   on a state-space formulation, hence allowing  for a quite flexible specification of structural components. 
A note in the documentation for \code{StructTS} alludes to the difficulty in performing optimization for even univariate structural time series models;
 this might explain the low number of \proglang{R} packages currently available for multivariate structural modeling. 
A crude search through \proglang{R} documentation using \code{RSiteSearch("structural time series")} \citep{graves2010sos} returns 114 documents matching our query. 
However, including the qualifier multivariate \code{RSiteSearch("multivariate structural time series")} returns 0 matches. 
 This supports the contention that   \pkg{Ecce Signum} can meaningfully add to R's ecosystem.

The other types of \proglang{R} packages that \pkg{Ecce Signum} overlaps with are those which perform signal extraction on high frequency time series. 
We use the term high frequency to mean series observed at the weekly, daily, or even higher frequencies. Here we discuss a few packages for performing high frequency seasonal adjustment.
The \pkg{forecast}  \proglang{R} package \citep{hyndman2008automatic, hyndman2020forecast}   incorporates multiple functions that can accomplish high frequency seasonal adjustment; \code{stlf()} uses an STL decomposition, \code{mstl()} is a variation for multiple seasonal periods, and \code{auto.arima()} can be utilized to perform dynamic harmonic regression  \citep{hyndman2018forecasting}.
  \pkg{forecast}   also includes an implementation of TBATS (Exponential Smoothing State Space Model With Box-Cox Transformation, ARMA Errors, Trend And Seasonal Components) \citep{delivera2011forecasting}.
\pkg{prophet}, a package developed at Facebook \citep{taylor2018forecasting}, attempts to provide a black box seasonal adjustment method focused upon forecasting. 
For well-behaved series this works well, but in our experience the many curious features in most daily time series still generate challenges for
 these sorts of automated methods.
\pkg{dsa}, a package developed for daily seasonal adjustment at Bundesbank \citep{ollech2018seasonal}, implements a version of STL.
A general overview of these methods can be found in \cite{ladiray2018seasonal}, and a comparison of different methods by their out-of-sample forecast accuracy can be found in \cite{timmermans2017cyclical}.
  
  The framework of \pkg{Ecce Signum} consists of an $N$-dimensional multivariate time series,
  which may be observed with missing values at various times for various components.  
  After the possible application of a log transform,  the model is written additively
  in terms of fixed effects (given through user-specified time series regressors) and
   latent stochastic processes (which the user specifies  during model declaration).
   The class of processes is fairly broad, including ARIMA and SARIMA models \citep{BoxJenkinsReinselLjung2015},
      as well as popular structural models discussed in \cite{Harvey1989} and \cite{HarveyTrimbur2003}.
 In broad strokes, the analyst proceeds from exploratory tools to a  putative model,
  which can then be fitted using Gaussian maximum likelihood estimation -- possibly accounting
   for additive outliers by entropy methods;  if there are latent stochastic processes present, 
    these are implicitly fitted as a result.   If the modeling is satisfactory,  the analyst can then
    proceed to applications such as forecasting or signal extraction.
    
    The paper is organized into the following sections:  Section 2 proceeds through the various stages
    of analysis that a user of \pkg{Ecce Signum} would utilize sequentially.   Then Section 3 provides
    more statistical detail about model parameterizations,  estimation methods, and vector embeddings.
    Several detailed examples are explored in Section 4;  these are vignettes in the package.  Appendix A 
    treats cyclical processes, and
    Appendix B
    discusses issues related
    to weekly time series.   Finally, Appendix C
    gives more granular detail on the functionality,  with a topology of the function
    call structure given through look-up tables.

\section{Stages of Analysis}

In the analysis of multivariate time series in \pkg{Ecce Signum}, there are several stages that are interdependent.  
 We discuss these in turn, in the sequence that they would typically be utilized in an application.

\subsection{Loading, Transforming, and Exploring}

We begin with entering the data  into an  \proglang{R} variable \textsl{data}, which is a $T \times N$-dimensional
  matrix involving $T$ time points and $N$ series.   If there are missing values, these can be encoded with ``NA,"
in which case the full sample size $T$ is defined by taking the size of the union of all temporal indices for each 
component time series.   A first stage of analysis is to call \code{sigex.load} with metadata such as the start date,
 period (i.e., number of seasons per year, if applicable), and epithets (i.e., shortened names or mnemonics) for
each component series. 

Secondly, a call to \code{sigex.prep} allows the analyst to apply a log transformation, aggregate component series
 (say if one needs to analyze the sum of all the series), restrict the temporal range,  select a subset of component series for
analysis, and plot.   The output we shall denote as \textsl{data.ts}, and for convenience the cross-sectional
  dimension will be denoted  $N$  (even if a subset of series was selected).   Note that this function call may also be utilized
after some exploratory analysis to identify transformation,  range, and subcollection;  see Chapter 1 of
\cite{McElroyPolitis2020}  for background. 

Optionally,  spectral plots can be viewed with a call by \code{sigex.specar},
 which  is just a wrapper for  the native \code{spec.ar} of \proglang{R}.  To the trained analyst, such plots provide insight
into the salient dynamics of each component time series, and can assist one in positing unit root model structures in 
subsequent analyses; see Chapter 9 of \cite{McElroyPolitis2020},
 or the discussion in \cite{McElroyRoy2021}.
   For instance, a sharp peak in the spectral density at a frequency $\lambda \in [0,\pi]$ indicates
 that a unit root of the form $e^{i \lambda}$ may be present in the autoregressive representation of the time series.

\subsection{Model Declaration}

Models in \pkg{Ecce Signum} have two facets: stochastic effects and fixed effects.  Fixed effects determine the mean
 of the time series through specified regressors, one for each component series.  Just as with RegARIMA models
  (Findley et al., 1998),
 once these fixed effects are subtracted out what remains is a mean zero stochastic process, which can be modeled
with one or more latent processes\footnote{ Called ``unobserved components" in the econometrics literature.
  We use the term latent process to avoid confusion with the components of a vector time series.}.
  Let $\{ X_t \}$ denote the transformed multivariate time series,  for which our sample corresponds to  \textsl{data.ts}.
  Our modeling framework is given by
\begin{equation}
\label{eq:reg-latent}
	X_t  =    z_t^{\prime} \, \beta +  Y_t .
\end{equation}
  	Here $\{ X_t \}$ is the data process, an $N$-dimensional time series  with time-varying mean 
 $\EE [ X_t ] = z_t^{\prime} \, \beta$ by our modeling assumption.   The stochastic process $\{ Y_t \}$
 has mean zero.     For each $t$ in the sample period $\{ 1, 2, \ldots, T \}$,  the regressors can be placed in 
 a matrix  $z_t$  of dimension $r \times N$, where $r$ represents the union of all fixed effects for all
component series.    Because  each component series has separate regression effects   (even if an identical
 regressor is used for several component series, the parameters will be distinct by assumption),
 the structure of $z_t$ is block diagonal.   In particular, if $z_t (j)$  is an $r_j$-dimensional vector of regressors
 for component series $j$ (for $1 \leq j \leq N$),  then
\[
    z_t =   \left[  \begin{array}{cccc}   z_t (1) &  0 &  \ldots &  0 \\   0 &  z_t (2) &  \ldots &  0 \\
	\vdots & \vdots & \vdots & \vdots \\   0 &  \ldots & 0 &   z_t (N)  \end{array} \right].
\]
 Then with $r = \sum_{j=1}^N r_j$,  we have
$\beta^{\prime} = [ {\beta (1) }^{\prime},   {\beta (2) }^{\prime}, \ldots, {\beta (N) }^{\prime} ]$
  and the fixed effect for the $j$th component series is $ { z_t (j) }^{\prime} \, \beta (j)$. 
 More details on adding fixed effects are given below,  and  in Section \ref{sec:reg-effects}.

 Our first step in modeling $\{ X_t \}$ is to build  the stochastic 
    model for $\{ Y_t \}$.  Here we use \textsl{mdl} as the variable (a list object) storing all
 the model information.  We initially  declare \textsl{mdl} to be NULL, and then use repeated calls of
 \code{sigex.add} to insert latent processes into the model.  The key inputs to these function calls
 are the scalar differencing operator $\delta (B)$
 (where  the backshift operator $B$  acts on a time series via $B \, X_t = X_{t-1}$),
   the rank configuration for the driving white noise,
 the time series model type (e.g., VARMA or SARMA, etc.) with  ancillary information such as the model order,
 and an epithet (e.g., trend, irregular, weekly seasonal, etc.).  If a component is stationary, simply input
 $\delta (B) = 1$; otherwise, input various polynomials as coefficient strings, taking care that each
 non-stationary latent process has a unique polynomial\footnote{Identification problems in signal extraction
  occur if diverse latent components share common unit roots; see \cite{bell1984} and \cite{McElroy2008} for
  discussion.}.   In particular, suppose there are $K$  vector latent processes $\{ S_t^{(k)} \}$  (for $1 \leq k \leq K$)
  that are related to the stochastic portion of $\{ X_t \}$ via  
\begin{equation}
 \label{eq:stoch-latent}
   Y_t = \sum_{k=1}^K  S_t^{(k)}.
\end{equation}
  For instance,  $\{ S_t^{(1)} \}$ might correspond to a trend, and $\{ S_t^{(2)} \}$ to a transient stationary structure.
  Each latent component has an associated (scalar) differencing operator $\delta^{(k)} (B)$ such that
\begin{equation}
\label{eq:diff-latent}
   \delta^{(k)} (B) S_t^{(k)} =  \underline{S}_t^{(k)},
\end{equation}
 where $ \{ \underline{S}_t^{(k)} \}$ is a mean zero, covariance stationary  vector stochastic process. 
 The differencing in (\ref{eq:diff-latent}) is applied in the same way to each component of the vector process.
 The stationary model structure can be specified through ARMA/SARMA (the same model is used for each
  component), or a VARMA/SVARMA, or a cyclical model.  Generally, all the models implemented in \pkg{Ecce Signum}
   are special cases of the causal moving average form
\begin{equation}
\label{eq:causal-ma}
  \underline{S}_t^{(k)} = \Psi^{(k)} (B)  \epsilon_t^{(k)},
\end{equation}
  where $\Psi^{(k)} (z)$ is a matrix power series.   The coefficient matrices are multiples of the identity in the case
   of ARMA/SARMA models, or a cycle model,  but for VARMA/SVARMA these matrices can be non-trivial;
    further details are in Section \ref{sec:serial-param}.
   The vector white noise process $\{ \epsilon_t^{(k)} \}$ is independent and identically distributed (i.i.d.) with unspecified marginal distribution,
   and covariance matrix $\Sigma^{(k)}$ that can be reduced rank; further details are in Section \ref{sec:cov-param}.

The specification of these components takes some care and experience.
  Note that any unit roots deemed to be present
in the data process' autoregressive representation must be featured among one of the latent processes,
 since the product of the differencing polynomials for the latent processes equals the differencing 
 polynomial for the data process \citep{HillmerTiao1982}.
  In particular, setting $\delta^{(-k)} (z) = \prod_{j \neq k} \delta^{(j)} (z)$, 
  it follows from (\ref{eq:stoch-latent}) and (\ref{eq:diff-latent}) that
\begin{equation}
\label{eq:diff-relations}
   \delta (B)  Y_t =  \sum_{k=1}^K   \delta^{(-k)} (B)  \underline{S}_t^{(k)},
\end{equation}
 where $\delta (z) =  \prod_{k=1}^K \delta^{(k)} (z)$.   
  The right hand side of (\ref{eq:diff-relations}) is a sum of stationary processes,  and hence $\delta (B) Y_t$ is
   stationary too,  and yet if there are any common unit roots among any of the $\delta^{(-k)} (z)$ then
   $\{ Y_t \}$ has been over-differenced -- this will generate estimation problems, because the resulting process
   will be non-invertible (Chapter 6 of \cite{McElroyPolitis2020}).  
   Therefore it is important to specify the
   various $\delta^{(k)} (z)$ such that they are relatively prime.  Henceforth
   $ \underline{Y}_t = \delta (B) Y_t$ defines the mean zero stationary process given by (\ref{eq:diff-relations}).

The second stage in model declaration is building in fixed effects; at a minimum, one must call 
 \code{sigex.meaninit}, which ensures that a polynomial trend is inserted with order corresponding
 to the number of specified trend differences in the first stage.   This is important, because 
 applying differencing to the data process will typically leave a non-zero constant mean, which is
 automatically associated with the leading coefficient of a polynomial trend effect.  Next, if additional
 fixed effects are desired (e.g., calendar effects, level shifts, etc.) then these can be incorporated
 through calls to \code{sigex.reg}.    This function requires that the regressors have already been
constructed for the full sample span $\{ 1, 2, \ldots, T \}$, so if a reduced range has been considered
 any regressors inputted to the program must be appropriately (and manually) culled.  
  Because each component series can have different regressors, one must set up a loop over the $N$ components
 to allocate the specified regressors.

For moving holiday effects such as Easter, the utility \code{gethol} for daily time series can be used.
 One inputs the desired calendar dates, and a window of days before and after each occurrence of the holiday when
it is deemed that economic activity will be increased relative to the rest of the year;
 this mimics  the Genhol routine \citep{Bureau2015}
  of X-13ARIMA-SEATS  for constructing moving holiday regressors.
  In particular, if $t$ is the index corresponding to a holiday, and $f$ and $a$ indicate the number of days before and
  after the holiday to be considered, then the window is $t-f, t-f+1, \ldots, t, \ldots, t+a-1, t+a$; the length of
  the window is $a +f + 1$. 
   A corresponding   sequence of regressors, computed using the start and end dates of the time series sample, is generated.

\subsection{Model Fitting}

By the fitting of a model, we mean that parameter estimates are obtained from the data.  \pkg{Ecce Signum} has
 two methods for doing this: Method of Moments (MOM) and  Maximum Likelihood (ML).
 Before these can be used,  initial values of the parameters need to be set up, and the practitioner makes choices
about whether a parameter is fixed or free.   A detailed discussion of parameterization and fitting is given in 
 Section \ref{sec:param}, and MOM is expounded in Section \ref{sec:mom}.

Typically, the ML method is preferred.   This is because the MOM method is limited to fairly simple latent process
 structures;  essentially,  the causal moving average portion $\Psi^{(k)}$ of the $k$th latent component model
 (\ref{eq:causal-ma}) is held fixed at an initial value specification,  and only $\Sigma^{(k)}$ is estimated.
 Moreover, the  MOM method is  not equipped to handle missing values,  and the estimation of 
 fixed effects is through OLS.  In contrast, ML allows for ragged edge missing values and  the most general types
  of model structures encoded in \pkg{Ecce Signum};  moreover,  the statistical estimates are efficient
   when the model is correctly specified and any non-Gaussian features are absent -- see pp. 52--59 of \cite{TaniguchiKakizawa2000},
  as well as the fourth appendix of \cite{HolanMcElroyWu2017} for details.
   The parameters for fixed effects correspond to Generalized Least Squares (GLS) estimates based upon
   the time series model parameters \citep{McElroyHolan2014}.
  Furthermore,  we can impose parameter constraints (Section  \ref{sec:constraint-param})  and generate
  test statistics for parameter significance.
  
  The demerits of ML are the larger computational resources required, as well as the usual pitfalls of nonlinear
   optimization.  Each likelihood evaluation has an expense relative to the size of $T$ and $N$, as well as 
    the model complexity  \citep{mcelroy2018recursive}.
     For models with many parameters, e.g., 100 or more,
    the parameter manifold is large and requires a time-consuming search.  As usual, there are no guarantees
    that the algorithm's termination corresponds to a global maximum.  Our approach in \pkg{Ecce Signum}
    is to re-parameterize constraints into an unconstrained (so-called)  pre-parameter space homeomorphic to Euclidean space,
as advocated by \cite{PinheiroBates1996}.
  Different initializations can be specified by the user,
 as well as what type of optimizer is desired;  the  pre-parameter corresponding to the  
  last successful  likelihood evaluation is saved to the global variables in the \proglang{R} workspace,
   so that if \textsl{optim} crashes the user can restart with a slight perturbation of this last value.
   
Upon termination of the optimization routine,  the user can examine the parameter values and assess the model
 fit.   This is possible by computing the time series  residuals and measuring their entropy via a Shapiro-Wilks normality test
 \citep{ShapiroWilk1965}  as well as Ljung-Box statistics
 \citep{LjungBox1978}.  
  More generally,  through modeling we are attempting to map the data vector to a maximum entropy residual;
   see Chapter 8 of \cite{McElroyPolitis2020}  for background on this modeling philosophy.  A serially uncorrelated Gaussian time series has maximum entropy,  and therefore the user can assess departures of the model residuals from this ideal target.
    
  More generally,  the transformed time series may have extreme values present -- see  \cite{McElroy2016measurement} for an overview
   of extremal time series, and the impact on signal extraction problems.  
   As discussed in   \cite{McElroyPenny2019},   non-Gaussian processes can be handled through non-linear
   forecasting and signal extraction algorithms, or by first applying an extreme-value adjustment procedure
     followed by a linear algorithm.   \pkg{Ecce Signum} follows the latter path, chiefly for simplicity and speed
     (non-linear methods typically require more nuanced modeling, and more expensive algorithms).
    This approach is in the tradition of intervention analysis:
    \cite{ChangTiaoChen1988} use an  additive outlier (AO) regressor
    \citep{Ljung1993},  and  \cite{Tsay1986}  considers more general types of outlier effects
    (e.g., level shifts).   
    
    We view the AO as an unknown stochastic additive effect, not a fixed effect;
    as shown in \cite{McElroyPenny2019}, entropy is maximized by deleting the corresponding observation.
    Hence, the treatment of AOs in \pkg{Ecce Signum} is by inserting an NA for any putative additive extreme,
    and comparing the likelihood.    The difference in log likelihoods has an interpretation as a test statistic
    for the significance of any collection of AOs.   The identification of extremes is an important facet of
     model building, and can be accomplished in tandem with fitting a specified model; as missing values are
     introduced for possible extremes, the remaining sample attains to more typically Gaussian characteristics,
     and the ragged edge Durbin-Levinson algorithms allow for quick computation.  

If multiple satisfactory models have been obtained,  we can perform  nested comparisons
using a difference of likelihoods \citep{McElroy2016nonnested}.
If particular pre-parameters are not significantly different
from zero,  they can be replaced by a zero (this is entered as a constraint) and the model re-fitted (as
 a nested model).   Another possible restriction is through the rank configuration of a latent process'
  white noise covariance matrix;  see \ref{sec:cov-param},  although in this case a statistical test
  arising from the difference of likelihoods is not valid (without a non-standard asymptotic theory)
  because the restriction lies on the boundary of the parameter space. 
  
    Through use of these tools,   we suppose the analyst has arrived at a final model, which is bundled
    using \code{sigex.bundle}:  this includes \textsl{data.ts} (where any detected AOs have been marked
    by inserting NAs), the model \textsl{mdl},  the fitted pre-parameters $\psi$,   and the transformation used.

\subsection{Applications of the Fitted Model}

 At this point in the analysis we can generate output from the fitted model for applications.
 We focus on casting (forecasting, aftcasting, and midcasting) and signal extraction.
 The function \code{sigex.midcast} generates midcasts for all missing values in the data,
 which are marked as usual with the NA symbol;    a number of forecasts and aftcasts to be generated as well. 
   \code{sigex.midcast} is a ``deluxe" version of the casting methodology, also generating the large
   covariance matrix between all casting errors; since this can be expensive to compute and store if there are hundreds
   of casts,  a faster version is available when uncertainty is not needed -- \code{sigex.cast} generates
   only forecasts and aftcasts, and presumes there are no missing values.

  Neither of these two function calls is complete without accounting
 for regression effects: both \code{sigex.cast} and \code{sigex.midcast} presume that fixed effects have
  already been removed from \textsl{data.ts},  and these regression effects need to be extended and added back
  to the series.   The reason this is not done automatically, is that  there is no way to know future values of
   general regressors encoded in \textsl{mdl};  clearly, polynomial and sinusoidal functions can be extrapolated/interpolated,
   but for holiday regressors or calendar effects a human is required to do the modifications.
   One approach to analysis is to load the data with extended regressors entered into \textsl{mdl}, and \textsl{data.ts} padded
   fore and aft with NA, up to the horizons that are anticipated to be needed. 
    
    For signal extraction, there are two main options: Wiener-Kolmogorov (WK) filtering or {\it ad hoc} (AC) filtering,
     as discussed in \cite{McElroyCasting}.
           In the first case, we have in mind a model that contains at least two latent
     processes,  and it is of interest to extract one of them (or possibly a sum of a subset of these latent processes).
     For the specified latencies, \pkg{Ecce Signum}  computes the minimum mean square linear estimate conditional 
     on the (ragged edge) data, along with time-varying uncertainty.   This uncertainty can be viewed as being composed
     of two portions: the so-called WK error and the casting error.
     In contrast, the AC case can be applied when there is only a single latency specified,  but the analyst must define
     and supply the desired (non-model based) filter.   Like the WK case, the optimal estimate is computed from the
      ragged edge data, but the time-varying uncertainty only depends on the casting error, as the WK error component
      is not defined.    In either case,  the extracted signal  $\{ U_t \}$ takes the form
\[
   U_t = \Upsilon (B) Y_t = \sum_{h = - \infty}^{\infty} \Upsilon_h Y_{t-h},
   \]
      where  $\Upsilon (B)$ is either a WK or AC filter.   In the WK case, $\{ U_t \}$ is viewed as an estimate of
      a stochastic process, such as a latent process $\{ S_t^{(k)} \}$ (or aggregation of such).  But for AC filtering,
      $\{ U_t \} $ is viewed as the actual target.
      
We begin with the WK case, and there are two methods available:  \code{sigex.signal} provides the exact filter
 matrix that yields the WK signal extraction at all time points (derived in \cite{mcelroy2015signal}), 
 but requires there be no missing values.
 (Although we can apply \code{sigex.signal} to ragged edge data that has been ``patched" with midcasts,
  we will not be able to account for midcasting uncertainty in the signal extraction uncertainty measure.)
  A secondary call to \code{sigex.extract} furnishes the actual extractions
    (along with point-wise confidence intervals defined as
   $\pm 2$ standard deviations, based on the extraction error covariance matrix), 
   but the signal extraction error covariances
  (assuming no missing values) as well as the filter weights are produced by \code{sigex.signal}.
  This method is not recommended for longer samples ($T > 500$), because a non-Toeplitz matrix of dimension 
   $NT$ needs to be inverted.

   The broader method \code{sigex.wkextract} is generally preferred, except in cases
   where the time-varying filter weights are needed.   The technique is described in \cite{McElroyCasting}, and involves
    first patching the series with midcasts -- and a set number of aftcasts and forecasts to handle the sample boundary --
    followed by filtering with the so-called WK filter.   This method also obtains both components of the signal
    extraction uncertainty,    accounting for missing values and the sample boundary.  
    However, unlike \code{sigex.signal} it assumes a
    truncation of the WK filter.   The truncation error
    can be reduced to any desired level by taking  lag cutoff to be  larger, but this induces a computation cost
    because  there is a corresponding increase to the number of  aftcasts and forecasts (with uncertainty) that are needed.

   An additional feature is the ability to generate casts of signals; automatically the signal is extracted for
   time points $\{ 1, 2, \ldots, T\}$, but for a specified horizon $H$  we can obtain forecasts
   for times $\{ T+1, \ldots, T+H \}$ and aftcasts for times $\{  1-H, \ldots, 0 \}$.   A second auxiliary feature
   is the extraction of linear combinations of a signal:   suppose that our  extraction is
\[
    \Xi  (B)  U_t  = \sum_{j=0}^p \Xi_j U_{t-j} 
\]
 for coefficient matrices $\Xi_j$.   The matrix polynomial $\Xi (z)$ is passed into \code{sigex.wkextract}, 
   and the extraction with appropriate covariance is returned.  Note that such extended
  extractions can also be obtained through use of \code{sigex.signal}, by applying appropriate linear combinations
   to both the filter matrix and the extraction error covariance matrix, but in \code{sigex.wkextract} the 
   error calculations are done using frequency domain instead of time domain.   Mathematical details of these matters 
   can be found in \cite{McElroyCasting}.   
  
   A second option is to generate AC extractions.  The desired filter $\Upsilon (z)$ is entered as an array object,
   in a truncated format:
\[
 \Upsilon (z) \approx \sum_{h = -c}^{m-1-c}  \Upsilon_h  z^h,
 \]
   where $m$ is the total number of nonzero coefficient matrices, and $c$ is a shift parameter that indicates the 
   proper centering.  For instance, setting $c = 0$ indicates that the filter is concurrent, i.e., the output depends
   only on present and past data.  Then \code{sigex.adhocextract} computes the filtered estimates,
   first patching and extending the series as needed; this routine functions much like \code{sigex.wkextract},
   only the AC filter is used instead of the WK filter.   
   Also, if we really desired a signal of the form $\Xi (B) U_t$,  this amounts to
   \[
     \Xi (B) U_t = \Xi (B) \Upsilon (B)  Y_t,
\]
 so our AC filter would just be redefined as $\Xi (B) \Upsilon (B)$.

 Plots of casts and extractions, with uncertainty indicated, is a final step in the analysis.
 Note that regression effects should be added back to casts, and may be added to certain extractions.
  For example, an Easter moving holiday regressor would be added to the final extracted seasonal component.
 To pair estimated parameters with the appropriate regressors, \code{sigex.fixed} can be used.
  For plotting,  \code{sigex.graph} can be used to overlay extracted signals if the data has already been 
 plotted;  this function presumes the format of the output of \code{sigex.extract}.   
   The second argument of \code{sigex.graph} accepts fixed effects that are added onto the extraction
  in the first argument.  The starting date and period are also entered, and it is possible 
  to vertically  displace plotted signals (if this is visually desirable).  For example, the seasonal component will be plotted
  around zero,  and hence may be too far below the data and trend extraction to be easily visible -- it can be shifted
  up in such a case.  
 The extracted signals can also be assessed as to whether they contain only the proper dynamics.  One can examine
  the spectral density through a call to \code{sigex.specar}, and of course the autocorrelations can be checked as usual.

    It may be of interest to examine the WK filter coefficients, as well as the frequency response function.
  \code{sigex.wk} can be used to compute and plot   the WK coefficients $\Upsilon_h$.   
    In particular, there are multiple panels and the $jk$th panel
  (for $1 \leq j,k \leq N$) plots the $jk$th entries of  $\Upsilon_h $ for $h = -m, \ldots, m$.
  For example, the weights on the second input series used to extract the first output series are given by $j=1$ and $k=2$.
  Similarly, \code{sigex.frf}  computes (and can be plotted in multiple panels
   by calling  \code{sigex.getfrf}) the  frequency response function 
  $\Upsilon (e^{-i \lambda})$.   The output is an array, with elements corresponding to a mesh of frequencies $\lambda$
  taken in the interval $[0,\pi]$.   Note that for purposes of inverting the frequency response function into coefficients,
  it is always sufficient to restrict to the interval $[0, \pi]$,  because
  \[
   \Upsilon_h = \frac{1}{  2 \pi} \int_{-\pi}^{\pi}  e^{i \lambda h} \Upsilon (e^{-i \lambda}) \, d\lambda
    =  \frac{1}{ \pi} \int_0^{\pi} \Re \left[ e^{i \lambda h} \Upsilon (e^{-i \lambda}) \right] \, d\lambda
\]
 by change of variable.    (This integration is discretized and calculated in \code{sigex.wk}.)

\section{Detailed Aspects}

\subsection{Model Parameterization}
\label{sec:param}

\pkg{Ecce Signum} makes use of unconstrained parameterization, which means there exists a bijection between all
 model parameters and Euclidean space.   Specifically, all model parameters are stored in a list object \textsl{param}
 (whose format depends on the specified model), and there exists a bijection that maps a real vector $\psi$ to
 \textsl{param}.  The vector $\psi$ is further decomposed into three portions:  $\xi$ (which maps to covariance matrices
 for driving white noise of the latent processes),  $\zeta$ (which maps to time series models, such as ARMA, for the 
latent processes), and $\beta$ (which maps to the  regression parameters for all component series).  
 The functions \code{sigex.par2psi} and \code{sigex.psi2par} pass \textsl{param} to $\psi$ and back.

  Why do we do this?   The format of \textsl{param} involves the Generalized Cholesky Decomposition (GCD)
   of covariance matrices as well as things such as
 AR coefficients, organized in a list that corresponds to component series;  hence a user can more easily interpret
 the quantities in \textsl{param}.   But $\psi$ consists of the unconstrained parameterization of covariance matrices
 and stable AR polynomials and so forth;  it is not easy to understand what a particular component entry of $\psi$
 corresponds to.    Typically the user will print \textsl{param}, which has a readable structure;
 the standard user will not need to understand the mapping of $\psi$ to \textsl{param}.

\subsubsection{Covariance Matrix Parameterization}
\label{sec:cov-param}

 First, the parameterization of a non-negative definite matrix  $\Sigma$ proceeds along the lines
  described in \cite{mcelroy2017multivariate}.
  Specifically, the GCD (described in \cite{GolubVanLoan2012}) decomposes a possibly singular $\Sigma$ into 
 unit lower triangular $L$ and diagonal $D$ such that
\begin{equation}
\label{eq:gcd}
  \Sigma  = L \, D \, L^{\prime}.
\end{equation}
  For an $N$-dimensional matrix, there are $\binom{N}{2}$ lower triangular entries in $L$ that can be any real number.
  The matrix $D$ has $N$ non-negative entries, which are all strictly positive if $\Sigma$ is positive definite (invertible).
  Hence the total number of parameters required is $\binom{N+1}{2}$.   If the rank of $\Sigma$ is less than $N$,
 then at least one entry of $D$ is zero, and the corresponding column of $L$ becomes obsolete in the decomposition.
To determine the requisite number of parameters,  one must further specify which entries of $D$ are zero; we call
 this a {\it rank configuration}.  Clearly, if $d_j$ (the $j$th diagonal entry of $D$) is zero, then $N - j$  free parameters
 in $L$ become obsolete and can be removed from the needed parameterization.  (In the case $j=N$, no
 reduction occurs.)   

The complement of the rank configuration corresponds to entries of $D$ that are positive, and the variable \textsl{vrank}
 captures this. 
  If the rank is reduced, the number of parameters needs to be computed, and \code{sigex.param2gcd} constructs an $L$
 matrix from a vector of real numbers, where any columns corresponding to the rank configuration are omitted
(i.e., $L$ is a rectangular matrix with potentially more rows than columns).   The positive entries of $D$ are parameterized
 by taking the exponential of a real number.   In this way $\xi$ is mapped into the various covariance matrices
 of \textsl{param}, one for each latent process.

\subsubsection{Serial Dependence Parameterization}
\label{sec:serial-param}

Second, the parameterization of the serial dependence in the stochastic process
 is accomplished through \code{sigex.par2zeta} and \code{sigex.zeta2par}.
 The exact mapping is dictated by the specified model structures.   For an ARMA model, both the AR and MA polynomials are stably
parameterized through the pacf discussed in \cite{McElroyPolitis2020}.     In particular,  for $j =1, 2, \ldots, p$  for a degree $p$ 
stable polynomial  (with a minus convention, so that $\phi (x) = 1 - \sum_{j=1}^p \phi_j z^j$)  we consider recursively 
parameterizing all polynomials of degree up to $p$, denoting the coefficients at the $j$th stage via
  $\phi_1^{(j)}, \ldots, \phi_j^{(j)}$.  
  For $p=1$, we have $\phi_1^{(1)} \in (-1,1)$, which is accomplished
 by taking $\phi_1^{(1)} = (\exp \{  \zeta_1 \}  - 1)/( \exp  \{  \zeta_1 \}  + 1)$.    Then for $2 \leq j \leq p$,  parameterize $\phi_j^{(j)}$ in the same
 way and update $\phi_1^{(j)}, \ldots, \phi_{j-1}^{(j)}$ via
\[
   \left[  \begin{array}{c}  \phi_1^{(j)} \\ \vdots \\ \phi_{j-1}^{(j)}  \end{array}  \right]
  =    \left[  \begin{array}{c}  \phi_1^{(j-1)} \\ \vdots \\ \phi_{j-1}^{(j-1)}  \end{array}  \right]
    -  \phi_j^{(j)}  \,    \left[  \begin{array}{c}  \phi_{j-1}^{(j-1)} \\ \vdots \\ \phi_{1}^{(1)}  \end{array}  \right].
\]
This  update can also be expressed as follows:  multiply the old coefficient vector by the matrix $A(j-1)$ that is given
 by the sum of a $(j-1)$-dimensional identity matrix and another matrix with $-\phi_j^{(j)}$ on the off-diagonal.  
 In this way we map real numbers $\zeta_1, \ldots, \zeta_p$ to $\phi_1^{(p)}, \ldots, \phi_p^{(p)}$.
 By applying the inverse of $A(j-1)$ recursively, we can also recover the original pre-parameters from a given stable 
polynomial.    For a SARMA model  (with seasonal period $s$),
\[
   \phi (B) \, \Phi (B^s) Y_t =   \theta (B) \, \Theta (B^s) \epsilon_t,
   \]
and  each polynomial   is stably parameterized.  There are four such polynomials --
 a regular AR polynomial $\phi$, a seasonal AR polynomial $\Phi$, a regular MA polynomial $\theta$, and a seasonal MA 
  polynomial $\Theta$,  all of which are defined using the minus convention of \cite{BoxJenkinsReinselLjung2015}.

For a VAR  model,  the parameterization of  
\cite{RoyMcElroyLinton} is used, 
 whereby real numbers are mapped into the space of stable matrix polynomials. 
  The functions \code{var2.par2pre} and \code{var2.pre2par} provide the mapping from $\phi (B)$ to the real numbers,
   and back again.       For VARMA and SVARMA, there are additional polynomials
 for the moving average and seasonal aspects.

Stochastic cycles come in four varieties.  The basic $n$th order Butterworth cycle is described in  \cite{HarveyTrimbur2003},
 and takes the form
\[
    {(1 - 2 \rho \, \cos (\omega) B + \rho^2 B^2) }^n  C_t = {(1 - \rho \, \cos (\omega) B)}^n  \epsilon_t
\]
 for a cyclical latent process $\{ C_t \}$, where $\{ \epsilon_t \}$ is white noise  (this can be multivariate). 
   Here $\rho$ corresponds to the persistency of
 the cycle, and is restricted to $(0,1)$, while $\omega$ is the frequency (or $2 \pi$ divided by the period of the cyclical effect).
 A higher order $n$ ensures that the high frequency aspects of the cycle are damped, so that the resulting stochastic process
 more purely represents only cyclical phenomena.   In the model specification, 
 the user specifies $n$, and can also input range parameters for $\rho$ and
 $\omega$.   This can be used to impose hard boundaries on the parameter values, although 
 \pkg{Ecce Signum}  considers an open interval
 dictated by the specified bounds and maps $\zeta$ to this interval using a logistic transform. 
   
   A second type of process  called the Balanced 
cycle  can also be used, and the spectral shape of the resulting process is slightly tilted -- as compared to the Butterworth cycle --
 in such a way that the high frequency content is lowered.   
 \cite{HarveyTrimbur2003} argue for the more favorable properties
 of the Balanced cycle, which can be specified in the same way as the Butterworth.   

 Finally, both cycles can be ``stabilized,"  which refers to a modification of the model that is non-invertible.    \cite{BellPugh1989} first experimented with such models, describing them as canonical component models.
 Any stationary stochastic process has spectral density $f$ that is non-negative, and if it is strictly positive we can consider
 subtracting off the minimum value (a positive number), thereby rendering $f$ shifted downwards such that it now
takes on a zero value.   A process having such a spectral density is said to be non-invertible
  (Chapter 6 of \cite{McElroyPolitis2020});
  while slightly problematic from the standpoint of fitting and forecasting,  such a process has the interpretative advantage
that it is maximally stable in the sense that as much white noise as possible has been removed.   The intuition here is
 that  we can decompose  a given stochastic process into the sum of a non-invertible process and an independent white noise
 process of variance equal to the minimum value of the spectral density -- an operation sometimes called ``canonization,"
 as it is at the core of the canonical decomposition routine of 
  \cite{HillmerTiao1982}.  More details on the cycle models are given in Appendix A.


The last model available is the ``damped trend."   This is essentially an AR(1) process where the user can place bounds
 on the range of the $\phi_1$ parameter, in the same manner as with the persistency parameter in the cycles.  Typically
 this specification would be used in conjunction with a trend differencing operator such as $\delta (B) = 1 - B$.  The resulting 
 model for a trend latent process  $\{ S_t \}$  would then be
\[
    (1-B)  (1- \phi_1 B)  S_t = \epsilon_t.
\]
 In  \cite{Trimbur2006} this type of trend interpolates  continuously between
   the random walk and integrated random walk trends, allowing one
 to more finely model and extract a trend with a varying degree of smoothness (higher values of $\phi_1$ correspond to 
 smoother extracted trends).  

\subsubsection{Regression Effects Parameterization}
\label{sec:reg-effects}

Third, the regression parameters for each series are given by $\beta$, the final portion of $\psi$.   This portion is
 straightforward; note that a mean effect for the differenced time series is always added through \code{sigex.meaninit},
 so there will always be at least $N$ parameters in $\beta$.   When a user decides to add a regression effect to 
 a component series,  the program checks to see whether this is in the null space of $\delta (B)$, the full differencing operator
 for $\{ Y_t \}$  described in (\ref{eq:diff-relations}).
 %
%
Consider applying $\delta (B)$ to the $j$th component  (for $1 \leq j \leq N$)  of  (\ref{eq:reg-latent}).   We obtain
\[
  \delta (B)   X_{tj}  =  \delta (B) {z_t (j) }^{\prime} \beta (j) + \underline{Y}_{tj},
\]
 where $\underline{Y}_{tj} $ is the $j$th component of $\underline{Y}_t = \delta (B) Y_t$.  
    As discussed below (\ref{eq:diff-relations}),  the process $\underline{Y}_t$ is invertible  so long as
    the various $\delta^{(k)} (z)$ are relatively prime.     If $\delta (B)$ completely annihilates $z_t (j)$,
 then there are no fixed effects at all;  because the fitting of non-stationary processes relies upon a factorization of
 the Gaussian likelihood that  features the  likelihood of the differenced data $\{ \underline{Y}_{t} \}$ 
  (see  \cite{McElroyMonsell2015} and \cite{McElroyCasting}),  it follows that any covariate with corresponding
 regressor  in the null space of $\delta (B)$  cannot be identified -- therefore in \pkg{Ecce Signum} 
   such a regressor is removed from  the model. 
      This is automatically done in \code{sigex.reg}, and the removal is done without further warning to the user.
 (The desired regression effect that the user enters is never incorporated into the model.)    For regressors not in the null space
 of the full differencing operator, $\delta (B) {z_t (j) }$ will be non-zero and hence generates the time-varying mean for
 $ \delta (B)   X_{tj} $.   This is accounted for in likelihood and casting calculations.

There is a type of weak converse to this discussion.   If we have included differencing in our model, without loss of 
 generality we can envision the inclusion of all regression effects that are in the null space of this $\delta (B)$,
 although such effects can never be identified.   For instance, if $\delta (B) = 1 - B^4$ then a basis of functions
 composing the null space consists of $1$, ${i^t}$, ${(-1)}^t$, and ${(-i)}^t$  with $i = \sqrt{-1}$. 
 This follows from the theory of ordinary difference equations, discussed in Chapter 5 of \cite{McElroyPolitis2020}.
    An individual latent
 process may contribute some (real) polynomial factors to $\delta (B)$.
  In the course of modeling a time series,  we may find that the innovation variance of one latent process is very close to zero
 (relative to the overall scale of the data).   In such a case, the stochastic process has  limiting form as the innovation
 variance tends to zero given by only the  fixed time-varying mean function in the null space;  in practice,
 we can replace such a latent process with the corresponding fixed effect.   

For example, consider a quarterly seasonal process with
 differencing operator $1 + B + B^2 + B^3$. 
 If this  seasonal process
 were found to have innovation variance close to zero, we could eliminate this from the model and instead enter
 a fixed effect corresponding to the null space of $1 + B + B^2 + B^3$, i.e.,  ${i^t}$, ${(-1)}^t$, and ${(-i)}^t$.
 Because it is awkward to do regression with complex numbers -- and using the fact that complex roots will always
 come in conjugate pairs if we are considering a real-coefficient polynomial -- we can instead work with cosines and sines.
 In particular,  if $e^{i \omega}$  (for $0 < \omega < \pi$) is a root of a differencing polynomial that is being eliminated from the model
due to low innovation variance, then $e^{-i \omega}$ is also a root  and a regression effect of the form
\[
   c \, e^{i \omega t} + \overline{c} \, e^{- i \omega t}
\]
 is in the null space, for a complex regression parameter $c$.  However, this can be rewritten as
\[
   \beta_1 \, \cos (\omega t) + \beta_2 \, \sin (\omega t)
\]
  with $\beta_1 = c + \overline{c}$ and $\beta_2 = i (c - \overline{c})$, which are both real parameters.  
 As a result, we can simply introduce the regressors $\cos (\omega t)$ and $\sin (\omega t)$ into
 the model.   If instead the root  of the differencing polynomial is $-1$, we introduce the regressor ${(-1)}^t$.

 We make a parenthetical remark here about levels and rates.  All regression effects are specified for levels (the original,
 undifferenced data process), and when moving to rates (or a  series  where $1-B$ has been applied) 
   the regressors are also differenced
 in a like manner, so the interpretation of the regression coefficients is unaltered.   Sometimes economists enter fixed effects
 into data that has already been differenced, but don't difference the regressors; then the resulting parameter estimates
 no longer correspond to that particular fixed effect as it is manifested in levels.   Since this is confusing (one can find transformations
 for the regression parameters), we avoid this and always specify the model for the time series (in levels) 
   before any differencing has been applied.

So far, we have discussed the incorporation of fixed effects that a user thinks may be appropriate.  Because a time series after
 applying full differencing will rarely have mean zero, even if it is stationary, it is prudent to include a constant regressor for each
 component in the equations for $\delta (B) X_{tj}$.   But what regression effect does this imply for the original series $\{ X_{tj} \}$?
 Again,  any regressors in the null space of $\delta (B)$ cannot be identified.   But we associate the constant regressor
 for the differenced series with a polynomial trend effect in the original series, which is explained as follows.
  We say a unit root of order $d$ is present in $\delta (B)$ if
 ${(1-B)}^d$ is a factor, where $ d \geq 0$.   The action of ${(1-B)}^d$ on a polynomial of order $d$  reduces it to a constant
  (Chapter 3 of \cite{McElroyPolitis2020}).  
  Hence we can insert the regressor $t^d$ in the model, and its coefficient is proportional to the mean effect.
 In particular, if $\delta (B) = {(1 - B)}^d \, \delta_S (B)$ is the factorization, then
\[
    \delta (B)  \,   [ \beta t^d ]  =   \delta_S (B) \,  d!  \beta = \delta_S (1) \, d! \beta.
\]
  The last equality follows from the fact that passing a constant through a filter yields the sum of the filter coefficients;
 $\delta_S (1) \neq 0$ because $1$ is not a root of $\delta_S (B)$, as all the unit roots were already accounted for in the
 factorization of $\delta (B)$.   Hence the mean is $\delta_S (1) \, d! \beta$,   and the coefficient of the trend polynomial
 equals the mean effect divided by $\delta_S (1) \, d!$.    In the case that $d=0$, \code{sigex.meaninit} allows the user
 to input a trend polynomial of any desired order, but if $d > 0$ then this input is ignored and only the regressor $t^d$ is
 incorporated.

\subsubsection{Fixed and Free Parameters}
 \label{sec:constraint-param}

It is possible to impose general linear constraints on the pre-parameters  $\psi$.  
 For example, one can fix specific elements of $\psi$ to a number, such as zero, or one can impose that
  a linear combination of components equals a constant.  We can enforce that two components of
  $\psi$ take the same value by imposing a linear constraint consisting of $1$ and $-1$ as the combination,
  and $0$ as the value that the linear combination is enforced to equal.   In general these constraints are encoded
  with the equation
  \begin{equation}
  \label{eq:psi-constraint}
   C \, \psi = b,
   \end{equation}
   where $b$ is a vector with length given by the number of constraints.  So each row of $C$ corresponds to one
   linear constraint, providing the linear combinations of $\psi$.   The row dimension of $C$ must be less than the 
    column dimension, and the difference between these is the number of free variables.   

   A limitation of this framework is that constraints are enforced on pre-parameters instead of the original,
   interpretable parameters; in applied work a researcher might be interested in enforcing a constraint on
   the parameters, but apart from the regression parameters it is difficult to see how elements of $\psi$ should
   be bounded to yield such a constraint.   Because the mapping of regression parameters to the pre-parameter
   is the identity (they are mapped to the subvector $\beta$ of $\psi$),  this case is easily handled.
   Otherwise, constraints for $\xi$ or $\zeta$ could arise from a prior fit of the model where it was subsequently
   determined that certain pre-parameters were not significantly different from zero.
    
    Here we determine     a real vector of free variables, called $\eta$, and a real vector of constrained variables $\nu$, such that
     together $\eta$ and $\nu$ make up the vector $\psi$.  Then likelihood evaluation is in terms of $\eta$,
     and this is the variable that is optimized over for maximum likelihood estimation.
     To determine $\eta$ and $\nu$ from a given $\psi$, and vice versa, consider the QR decomposition \citep{GolubVanLoan2012}
     of the matrix $C$:
\[
  C = Q \, R \, \Pi,
  \]
   where $Q$ is an orthogonal matrix, $R$ is unit upper triangular (with dimensions the same as those of $C$),
    and $\Pi$ is a permutation matrix.   Let $R = [ R_1 \vert R_2 ]$ be partitioned such that $R_1$ is square;
    it is invertible because it is upper triangular with ones on the diagonal.   Then we define $\eta$ to be
    the bottom portion of $\Pi \, \psi$, and we solve for $\nu$ in terms of $\eta$ as follows:
\begin{equation}
\label{eq:eta2nu}
  \nu  = R_1^{-1} \, Q^{-1} \, b - R_1^{-1} \, R_2 \, \eta.
  \end{equation}
   Finally, $\psi$ is constructed from $\eta$ and $\nu$ via
   \begin{equation}
   \label{eq:eta2psi}
    \psi  = \Pi^{-1} \, \left[ \begin{array}{c}   \nu \\ \eta \end{array} \right].
    \end{equation}
    It follows that (using $1$ and $0$ to denote either an identity matrix or
     a zero matrix of appropriate dimension)
    \begin{equation}
    \label{eq:eta2psi-final}
    \psi =  \Pi^{-1} \, \left(  \left[ \begin{array}{c}   0 \\   1 \end{array} \right] 
    	-    \left[ \begin{array}{c}   1 \\  0  \end{array} \right] \, R_1^{-1} \, R_2   \right) \, \eta
    		+ \Pi^{-1} \, \left[ \begin{array}{c}   1 \\   0  \end{array} \right] \, R_1^{-1} \, Q^{-1} \, b,
    \end{equation}
    which expresses $\psi$ as $A \, \eta$ plus a constant vector.
    These calculations are  performed by \code{sigex.eta2psi}, and \code{sigex.psi2eta} produces $\eta$ and $\nu$ from
     a given $\psi$.     Default values of \textsl{param}  are determined through
  \code{sigex.default}, which occurs by setting  $\eta$ equal to the zero vector and computing 
  $\psi$ via  (\ref{eq:eta2nu}) and (\ref{eq:eta2psi}).   This requires the matrix \textsl{constraint} to be defined,
  which has the format $[ b \vert C ]$.  This can be set to NULL if there are no constraints.  
  
 The model can then be fitted using \code{sigex.mlefit} and the output stored to a variable \textsl{fit}.
 Setting the last argument to ``bfgs"  tells the computer to use the
  BFGS  \citep{GolubVanLoan2012} algorithm for nonlinear optimization of the Gaussian likelihood,
 based on initial values dictated by \textsl{param}, but fixing parameters according to \textsl{constraint}, and an overall
 model structure given by \textsl{mdl}.   
  However,   we begin by obtaining $\psi$ from \textsl{param}, 
  and $\psi$ should satisfy (\ref{eq:psi-constraint}); the code  checks this condition.   
 At the end of fitting a model, we will typically have   the MLE for $\eta$ (denoted \textsl{eta.mle}).
   Then a call to \code{sigex.eta2psi} will yield \textsl{psi.mle}, and we can apply \code{sigex.psi2par}   to obtain
   \textsl{par.mle}.

  One can also  extract the Hessian  matrix if BFGS was used for model fitting.    Note that uncertainty for
 the model parameters only pertains to  $\eta$, so the dimension of \textsl{hess} is given accordingly.
  The inverse of \textsl{hess} is the asymptotic covariance matrix $V$ of the MLEs for $\eta$;   to get the 
   asymptotic covariance matrix for $\psi$, we compute 
   $A \, V  A^{\prime}$, where $A$ is the
 matrix described below (\ref{eq:eta2psi-final}).

\subsection{Method of Moments}
\label{sec:mom}

The goal of the MOM procedure is to obtain quick, rough estimates that might be used as initial values in
 the ML  routine,  or alternatively as final estimates for higher-dimensional time series data where
 ML  is not computationally feasible.
Background for MOM for latent process models is given in 
 \cite{McElroyRoy2018frobnorm}, although some more specific
 implementation details are discussed here.  
 If  $\Psi^{(k)} (z)$ in   (\ref{eq:causal-ma}) is known,  then by applying (\ref{eq:diff-relations})
 we can express the autocovariances   of $\{ \underline{Y}_t \}$ in terms of certain known
 matrix polynomials and the covariance matrices $\Sigma^{(k)}$;  by substituting sample
 autocovariances,  we can solve for estimators of the covariance matrices.   These are called the MOM
 estimates.


In practice several provisos mitigate the usefulness of this device.  First, the presence of fixed effects interferes with
 the population equations.  Second, the solutions to the equations yield symmetric matrices, but these are not
 guaranteed to be positive definite (pd).  Third, there may be model parameters involved in the $\Psi^{(k)} (z)$.
Fourth, missing values in the data generate obstacles to constructing the sample autocovariances.   

 To handle the first obstacle, we begin with a simple OLS in \code{sigex.momfit}, which is fast to compute and consistent
 notwithstanding serial correlation present in $\{ Y_t \}$.   Subtracting off the fitted regression effects, we proceed
 and now presume the de-meaned series has no fixed effects present at all.   We can then apply the MOM equations
 so long as there are no missing values (the routine will not work if there are NAs); if parameters $\zeta$ are present
 and enter into the $\Psi^{(k)} (z)$,  these are fixed at the values dictated by the default
  settings.  

The second obstacle is important to rectify.  In \pkg{Ecce Signum}  a covariance matrix can be assessed in terms of condition
 numbers discussed in \cite{mcelroy2017multivariate},
 which are direct functions of the diagonal entries of $D$ in (\ref{eq:gcd}).  
 In fact, it can be shown that the condition numbers equal $D_{jj}/ \Sigma_{jj}$, and they are typically viewed
 in log scale.  After taking the logarithm, a value of $- \infty$ corresponds to a singularity in $\Sigma$ that occurs
 at a particular index in the rank configuration  (certain partial correlations of $\Sigma$  are equal to $\pm 1$ in this case).    Moreover,  negative values of large magnitude (in log scale) indicate the matrix is
 close to singularity;  it is possible to alter the condition numbers to positive quantities such that the entries of $D$
 are altered and the whole matrix becomes better conditioned.   This can be done without altering the various partial correlations
 embedded in $L$, and is implemented in \code{sigex.renderpd}, as shown below.
 
    Let a $j$-dimensional covariance matrix
 be denoted $\Sigma_j$, which can be decomposed as $\Sigma_j = L_j \, D_j \, L_j^{\prime}$ for a unit lower-triangular
 $L_j$ and a diagonal matrix $D_j$.  Letting the lower row of $L_j$ be denoted $[\ell_j^{\prime}, 1]$, it is known 
  \citep{mcelroy2017multivariate}  that
\[
 g_{j+1} = d_{j+1}  +  {\ell}_{j+1}^{\prime} \, D_j \,  {\ell}_{j+1},
\]
  where $g_{j+1}$ denotes the lower right entry of $\Sigma_{j+1}$.  
Moreover, the $(j+1)$th condition number can be expressed  as
\[
  \tau_{j+1} = - \log \left( g_{j+1} / d_{j+1} \right) = - \log \left( 1+  {\ell}_{j+1}^{\prime} \, 
	D_j \,  {\ell}_{j+1}/ d_{j+1} \right).
\]
 These are recursive relations, so that for a given $N$-dimensional $\Sigma$ we can examine the upper left submatrix
 $\Sigma_j$ for $1 \leq j \leq N$ and recursively compute the condition numbers $\tau_{j}$.  
 Hence, if we wish to modify a given $\Sigma$ such that the condition numbers exceed
  some minimum threshold, then we must modify either $\ell_{j+1}$ or $D_j$.   Whereas the former is related to linkages between
 variables,   each $d_j$ is a partial variance (of the $j$th variable given the preceding variables $1$ through $j-1$);
 choosing to modify these to some new $\widetilde{d}_j$, we can see how the condition number changes accordingly.
 Thus, if we set $\tau_{j+1}$ equal to a desired threshold $\alpha$, and given that we have already modified
 in a recursive fashion the preceding $D_j$ to $\widetilde{D}_j$, we have the formula
\begin{equation}
 \label{eq:new.d}
  \widetilde{d}_{j+1} = { \left( \frac{ \exp \{ - \alpha \} - 1 }{ {\ell}_{j+1}^{\prime} \, \widetilde{D}_j \,  {\ell}_{j+1} }
	\right) }^{-1}.
\end{equation}
This suggests the following algorithm to produce a positive definite modification of a given $\Sigma$ matrix: compute the GCD,
 and iteratively check whether $1/d_{j+1} < (\exp \{ - \alpha \} - 1  )/ {\ell}_{j+1}^{\prime} \, \widetilde{D}_j \,  {\ell}_{j+1}$;
 if so, we replace $d_{j+1}$ by $\widetilde{d}_{j+1}$ in  (\ref{eq:new.d}) and increment $j$.  This yields a new 
\[
 \widetilde{\Sigma} = L \, \widetilde{D} \, L^{\prime},
\]
 where all the condition numbers are bounded below by $\alpha$.  In practice, the thresholds for $\alpha$ can be set according
 to the criteria discussed in  \cite{mcelroy2017multivariate}.  For instance, $\alpha = -1.66$ corresponds to a partial correlation of $.90$; this is 
 sufficiently distinct from unity, so as to ensure there is no signal leakage.
 Calling \code{sigex.reduce} with a specified  value of $\alpha$  (denoted by \textsl{thresh} in the code)
  replaces low logged condition numbers with \textsl{thresh},  rendering the matrix pd
 through \code{sigex.renderpd}.   

However, this discussion pertains to matrices that are non-negative definite.  The covariance matrices estimated by
 MOM can potentially have both positive and negative eigenvalues.   In this case a second usage of \code{sigex.reduce}
 occurs (setting \textsl{modelFlag} equal to \textsl{TRUE}) whereby logged condition numbers lower than \textsl{thresh}
 result in the corresponding component index being added to the rank configuration.  In other words, a low condition number
 for component $j$ will result in $j$ being added to the rank configuration.   Now the GCD algorithm will still run on a
 symmetric matrix where the entries of $D$ are possibly zero or negative; in such a case \code{sigex.reduce} 
 will automatically assign a non-positive value of $D$ to being below the threshold, and hence the rank configuration will
 be incremented.  The outcome for MOM is that matrices that are not estimated as being pd get replaced by reduced
 rank approximations -- essentially any negative values of the diagonal of $D$ get replaced by a zero.   

Once a reduced rank pd covariance matrix has been obtained,  one can then determine \textsl{param} accordingly.
 However, the process of reducing the rank changes the model structure, and hence \code{sigex.reduce} updates
 \textsl{mdl}.  This is a side-effect of MOM estimation, and may be undesirable in some applications (say, where we wish
 to keep all the covariance matrices as being full rank initially).

\subsection{Casting and Fixed Effects}
 
 The basic model structure (\ref{eq:reg-latent}) yields a $\{ Y_t \} $ time series that is mean zero.
   Optimal extraction means that we should filter $\{ Y_t \}$ after having removed fixed effects,
   and then add back those fixed effects that pertain to the particular signal.  However,
   in practice we do not know $\beta$ exactly,  and we instead subtract $z_t^{\prime} \widehat{\beta}$
   from the data to get the de-meaned data:
   \[
   \widehat{Y}_t =  X_t - z_t^{\prime} \widehat{\beta}.
   \]
   If there are missing values, or extreme-value adjustment is being performed for an AO,
   then there is already an NA in $\widehat{Y}_t$,  and this will be estimated with a midcast.
   Importantly, the casting algorithms are predicated on the time series having mean zero,
   which is approximately true of $\widehat{Y}_t$.   Let $\widetilde{Y}_t $ denote the modification 
   arising from casting, i.e., imputing missing values and adjusting extremes.  The discrepancy between
     $\widetilde{Y}_t $  and  $\widehat{Y}_t$ is
\[
  \varepsilon_t  = \widehat{Y}_t - \widetilde{Y}_t,
  \]
   which is also the casting error.   It is important to keep track of this quantity, because in many applications
   of signal extraction we want all the extracted latencies to exactly add back up again to the original data.
   The form this typically takes is the following:
\begin{align*}
 X_t & =  \widehat{Y}_t + z_t^{\prime} \widehat{\beta}\\
 	& =  \widetilde{Y}_t + \varepsilon_t + z_t^{\prime} \widehat{\beta} \\
 	& = \Upsilon (B) \widetilde{Y}_t  +  (1 - \Upsilon (B))  \widetilde{Y}_t + \varepsilon_t + z_t^{\prime} \widehat{\beta}.
\end{align*}
 In the first line we break out the estimated fixed effects.  In the second line we correct for missing values and
  extreme values.   In the last line we apply two complementary filters -- such as seasonal adjustment and seasonal
   filtering.   The casting error must be added to one of these two extractions  in order that all sums back to $X_t$.
   The fixed effects might be split up between the extractions.   For example, if $\Upsilon (B)$ is a seasonal adjustment
    filter (so that $1 - \Upsilon (B)$ is a seasonal filter),  then the casting error for extreme-value adjustment should
     be added to the seasonal adjustment (because it corresponds to an outlier, and is therefore not seasonal);
     fixed effects that correspond to moving holidays can be added to the seasonal component, but the trend polynomial
should be combined with the seasonal adjustment.   

The only case where this breaks down is if the original date is missing,
 i.e., $X_t$ is NA.   In such a case,  we replace the left hand side with its imputation -- essentially by
  subtracting the casting error from both sides of the equation, along with the fixed effects.
  For instance, if $X_t$ is missing but we wish to publish a seasonal adjustment, the sum of the
   adjustment and seasonal factor equals the imputation of $X_t$ given by $\widetilde{Y}_t + z_t^{\prime} \widehat{\beta}$.

\subsection{Embedding}
\label{sec:embed}

A vector embedding of a univariate process consists of rewriting the process as an $N$-variate time series
 of lower sampling frequency, such that the new sampling frequency equals the original sampling frequency divided by $N$.
  Sometimes embedding facilitates the modeling of a scalar time series.  For example, a univariate daily time series
   might be embedded as a weekly time series with $N = 7$,  so that day-of-week specific effects can be modeled
    in the mean or variance directly.   The function \code{sigex.daily2weekly}  provides such an embedding 
     given a specification of the day-of-week corresponding to the first component in the vector, and adds missing
     values to pad out the possibly incomplete first and last weeks.  An inverse is provided by \code{sigex.weekly2daily},
     which can be used on extracted components obtained from the weekly series.
     
\subsubsection{Embedding Filters}     
     
      Scalar filters can also be embedded,  turning them into a multivariate filter with Toeplitz structure.
      We can also apply embedding results to the moving average representation of a scalar
  series, allowing us to re-express it as a vector moving average.   
  %
 %
 We can embed a given high-frequency scalar
   process $\{ X_t \}$ as an $s$-variate  (where $s$ is a  positive integer)  low-frequency process   $\{ \xx_n \}$, where
the $j$th component  (for $1 \leq j \leq s$)  of ${\bf X}_n$ is $X_{ns+j}$. 
   This $j$th component is denoted as the $j$th {\it season's series},
    and the vector time series is  the {\it seasonal vector series}, which is denoted with a bold font.
    \cite{Gladyshev1961}, \cite{TiaoGrupe1980},
    \cite{Osborn1991}, and \cite{Franses1994} 
       have studied this type of embedding.  

  Define the backshift operator $B$, which acts on the high-frequency time index,  via $B \, X_t = X_{t-1}$.
   Similarly, the low-frequency lag operator is defined to be $L= B^s$, so that $L \, \xx_n = \xx_{n-1}$.  
 The $s \times s$ identity matrix is denoted $I_s$.
  Given a scalar Laurent  series  $\upsilon (z)$, its embedded matrix 
    Laurent series (cf. equation (2.7) of  \cite{TiaoGrupe1980}) 
    is defined to be the $s \times s$-dimensional matrix Laurent series 
  $ \Upsilon (L) = \sum_h  \Upsilon_h L^h$, where 
 \begin{equation}
 \label{eq:embed-filter-index}
  \mbox{the $j,k$ coefficients of}  \; \Upsilon_h \; \mbox{are given by} \; \upsilon_{j-k + sh}.
  \end{equation}
   It follows that  if $\upsilon (B)$ is a polynomial of degree $q$, then the degree
    of the matrix polynomial $\Upsilon (L)$  is $\lfloor q/s \rfloor  + 1$.
  Consider the following relations of filter input to filter output
   (expressed as operating on the de-meaned process):
  \begin{align*}
    U_t & = \upsilon (B) \, Y_t \\
    \uu_n & = \Upsilon (L) \, \yy_n.
   \end{align*}
   We verify that these two equations are equivalent, if we utilize the rule (\ref{eq:embed-filter-index}).
 The $j$th component (for $1 \leq j \leq s$) of the second equation (with $e_j$ the $j$th unit vector) indicates
 \[
   \uu_{n,j} =  e_j^{\prime} \, \sum_{h \in \ZZ} \Upsilon_h \, \yy_{n-h} 
     = \sum_{h \in \ZZ}  \sum_{k=1}^s \upsilon_{j - k + s h} \, Y_{ (n-h)s  +k }
     = \sum_{ \ell \in \ZZ}  \upsilon_{\ell}  \, Y_{ns +j - \ell} = \upsilon (B) Y_t,
    \]
    which uses the change of variable $\ell = j-k + s*h$.  
    
     Next, consider the case of a finite filter
    $\upsilon (B)$ that has a total of $m$ coefficients, and such that $\upsilon_0$ is the $(c+1)$th coefficient,
     where $0 \leq c < m$ is the integer shift parameter.   (In the documentation for
      \code{sigex.hi2low} and \code{sigex.adhocextract}, $L$ is used for the filter length.)
            This means that the non-zero coefficients
     are $\upsilon_{-c}, \upsilon_{1-c}, \ldots, \upsilon_0, \ldots, \upsilon_{m-1-c}$.  So $c=0$ corresponds to a
      causal filter, where only present and past observations get weighted; if $m$ is odd and $c = (m-1)/2$,
      then the filter weights an equal number of past and future observations.   The function
       \code{sigex.hi2low} calculates the corresponding non-zero filter matrices $\Upsilon_h$,
       and determines the appropriate shift parameter for the low-frequency multivariate filter.
       From (\ref{eq:embed-filter-index}), we have
\begin{equation*}
  \ldots, \;   
        \Upsilon_{-1}  = \left[ \begin{array}{ccc} \upsilon_{-s} & \ldots & \upsilon_{1-2s} \\  \vdots & \ddots & \vdots \\
         \upsilon_{-1} & \ldots & \upsilon_{-s}    \end{array} \right], 
   \;
   \Upsilon_{0} = \left[ \begin{array}{ccc} \upsilon_0 & \ldots & \upsilon_{1-s} \\ 
    \vdots & \ddots & \vdots \\ \upsilon_{s-1} & \ldots & \upsilon_0
   \end{array} \right], \;
  \Upsilon_{1}  = \left[ \begin{array}{ccc} \upsilon_s & \ldots & \upsilon_1 \\  \vdots & \ddots & \vdots \\
   \upsilon_{2s-1} & \ldots & \upsilon_s 
   \end{array} \right],   \; \ldots 
\end{equation*}
   We need to determine the length $M$ of the multivariate filter, and its shift parameter $C$.   Clearly,
   if any $\Upsilon_h$ has at least one non-zero entry then it must be included in the array object for the multivariate
    filter.   We begin by padding $\upsilon_{-c},   \ldots, \upsilon_{m-1-c}$ with $\ell$ zeroes on the left and $r$ zeroes on
    the right, such that: (i) $\ell + c \equiv 0 \, \mbox{mod} \, s$ and $0 \leq \ell < s$, (ii) $m+ \ell + r \equiv 0 \, \mbox{mod} \, s$,
     and such that $r$ is the smallest integer $ \geq s$ with this property. 
     
      First, (i) guarantees that the number of coefficients $\upsilon_k$ with
      $k < 0$ is divisible by $s$, and is the smallest such number.  Examining the bottom ($s$th) row of
      the $\Upsilon_h$ matrix,  we have the values $\upsilon_{ sh +s -1}, \ldots, \upsilon_{ sh }$,  
       so if all $\upsilon_k = 0$ for $k < - (\ell + c)$, then $\Upsilon_h = 0$ if $h < - (\ell + c)/s$.
       This means that the non-zero matrices are $\Upsilon_{ - (\ell+c)/s}, \ldots, \Upsilon_{ M-1 - (\ell + c)/s}$, with $M$ to be
        determined.  Hence the low-frequency shift parameter is $C = (\ell + c)/s$.    

Second, (ii) ensures that the entire length of the padded scalar filter is divisible by $s$, given that with the padding from
 (i) the filter is currently of length $\ell + m$.   We need to add at least $s$ zeroes on in this stage so that the
  bottom row of $\Upsilon_h$ with highest index is   zero except possibly for the bottom right entry $\upsilon_{ sh}$;
   if $\upsilon_{sh+1}$ were non-zero,  then it would appear in $\Upsilon_{h-1}$, indicating that the matrix is non-zero.
   It follows that $M  = (m+ \ell + r)/s$.

\subsubsection{Embedding From Daily to Weekly}

  We discuss some of the details on the weekly embedding.   Dates are stored in a month-day-year format
   as a three-component vector, and the utility \code{date2day} computes the day index (between $1$ and $366$)
   corresponding to any such date.  The inverse of this routine is \code{day2date}, and \code{day2week} 
   yields the day-of-week ($1$ for Sunday, $2$ for Saturday, etc.) corresponding to a given calendar date.
   To compute the week index corresponding to a given date, \code{sigex.daily2weekly} uses the following procedure:
   the day index is first computed from the date, and we determine the day index corresponding to 
    the first day of the first week.  Specifically, between $0$ and $6$ NAs are pre-pended to form the first week,
    with this number of NAs determined by what day-of-week corresponds to the calendar date, and 
    what type of day corresponds to the first component (this is called \textsl{first.day}, and is chosen by the user).
    Let $\ell$ denote the day index for the first day of the first week in the embedded series;
    note that $\ell$ can be negative (e.g., the calendar date is very early in the year, and we pad with enough NAs).
    Then the day indices for that first week are $\{ \ell, \ell+1, \ldots, \ell + 6 \}$.
     Our rule is that these indices corresponds to week $w$ of that year if and only if
     $ 7 (w-1) + 1 \in \{ \ell, \ell+1, \ldots, \ell + 6 \}$.   The logic is that if  $1 \in \{ \ell, \ell+1, \ldots, \ell + 6 \}$
     then this must be the first week (as it includes the first day index), and hence $w = 1$; thus the formula
      follows.   An equivalent condition is that
      \[
       w =  \lceil (\ell - 1)/7 \rceil + 1.
\]
 The lowest value that $\ell$ can take is $-5$, in the case that $6$ NAs were padded on and the calendar date is
  the first day of the year;  in that case $w = 1$, so we always get $1 \leq w \leq 53$.   For example,
   January 1, 2020 was a Wednesday.   Suppose we wish to have Saturday as our first series, but our time
    series has a start date of February 28, 2020.   Hence, there will be $6$ NAs added onto the first week;
    the day index for the start date is $59$, so $\ell = 53$ and $w = 9$.    The first ``custom" week covers
     December 28, 2019 through January 3, 2020,  and the ninth custom week includes February 22, 2020
      through February 28, 2020.  Note that week one will typically have a few days from the prior year; here
       the start year for the weekly series is 2020, even though four days of the first week are from 2019.
    
    When passing back to a daily series via \code{sigex.weekly2daily},  the original start date can be recovered
    as follows.   Compute the difference between the day-of-week of  January 1 of the start year (for the weekly series)
    and the given \textsl{first.day}, and add seven if this difference is negative; the result, called 
    \textsl{day.lead},  gives the number of days prior to January 1 that are included in the first week, and hence
\[
 \ell  = 7 \, (w-1) - \mbox{ \textsl{day.lead} } +1.
 \]
    From this day index $\ell$, we can determine the corresponding start date via \code{day2date}; however,
     if $\ell \leq 0$ (this can happen if $w=1$) then we need to decrement the year by one and add to $\ell$ the day index
     of December 31 for the prior year (i.e., either $365$ or $366$).   If we started by embedding a daily time
      series of a certain start date, use \code{sigex.daily2weekly}, and then invert with \code{sigex.weekly2daily},
      we will not recover the exact start date unless \textsl{first.day} corresponds to the day-of-week of
      that calendar date -- because otherwise, additional NAs are added and the daily series gets lengthened
      (so that the total sample size is a multiple of $7$).

\section{Illustrations}

We consider several numerical illustrations of the capabilities.   These are meant to be consecutive,
 in the sense that commands documented in a prior example are not explicitly re-visited in subsequent
  examples.

\subsection{Non-Defense Capital Goods}
\label{sec:ndc}
 
We first examine    data for   monthly shipments and new orders from the 
Manufacturing, Shipments, and Inventories survey,\footnote{Seasonally adjusted monthly data covering 
 the sample period is January 1992 to April 2020,  downloaded July 21, 2020 (4:45 PM),
  U.S.  Census Bureau, obtained from
 \url{https://www.census.gov/mtis/index.html} by selecting 
  Non-Defense Capital Goods,  and either Value of Shipments or  New Orders.} or
  \textit{ndc} for short.
  The data for New Orders is not available at January 1992, since this series starts at February 1992 --
  so this value is entered as an NA.   This is an example of ragged edge data.

 We perform a bivariate analysis by fitting a VAR model.
   	 After loading the data into a variable named \textsl{ndc}, 	 preliminary metadata is assigned as follows:
\begin{verbatim}  	 
dataALL.ts <- sigex.load(data       = ndc, 
                         start.date = c(1992, 1),
                         period     = 12,
                         epithets   = c("Shipments", "NewOrders"),
                         plot       = TRUE )
\end{verbatim}  
 These commands attach a starting date and frequency, as well as names for the series.
 The last argument of \code{sigex.load} generates a time series plot.
  On the basis of this plot, we decide (below) to utilize no transformation for the data.
  
 Next, data transformations and subsets of series can be selected.  Here, we will examine both series,
  so \textsl{subseries} is set equal to \verb!c(1,2)!.   The Boolean \textsl{aggregate} can be used
  to define a cross-sectional sum corresponding to \textsl{subseries}, but in this example that option is
  not utilized.  The \textsl{range} argument can be used to select a window of times, but is
  set to \verb!NULL! if the full span of the data is to be used.
 \begin{verbatim}
data.ts <- sigex.prep(data.ts.  = dataALL.ts,
                      transform = "none",
                      aggregate = FALSE,
                      subseries = 1:2,
                      range     = NULL,
                      plot      = TRUE )
\end{verbatim}
 Next, we move to model declaration, first recording the dimensions via
\begin{verbatim} 
N <- dim(data.ts)[2]
T <- dim(data.ts)[1]
\end{verbatim}
  We will fit a VAR model to the differenced data, and we use Yule-Walker estimation to
  identify the model order $p$ and get preliminary estimates.
\begin{verbatim}
ar.fit    <- ar.yw(diff(ts(ndc[2:T, ])))
p.order   <- ar.fit$order
par.yw    <- aperm(ar.fit$ar, c(2, 3, 1))
covmat.yw <- getGCD(ar.fit$var.pred, 2)
var.out   <- var.par2pre(par.yw)
psi.init  <- as.vector(c(covmat.yw[[1]][2, 1], log(covmat.yw[[2]]),
                        var.out, colMeans(diff(ts(ndc[2:T, ])))) )
\end{verbatim}

 The first line fits a VAR model using the native \proglang{R} function \code{ar.yw}.   Note that we exclude
  the first observation because it is missing for the NewOrders series.  We also apply differencing, as there
  seems to be some change in level for the series.   The second line extracts the identified model order ($p=7$ in
   this case), and the third line stores the fitted parameter in an array of dimension $2 \times 2 \times 7$.  
   The fourth line takes the estimated innovation covariance matrix, and produces the GCD via \code{getGCD}.
   The output is a list object, where the lower triangular entry of $L$ and the log of the diagonal entries of $D$
   are entered into the appropriate positions of \textsl{psi.init} in line seven.   Line five maps the array of VAR
   parameters via \textsl{var.par2pre} to a pre-parameter vector  \textsl{var.out}   of length 28.
     Lines six and seven assemble \textsl{psi.init} from the 
   Yule-Walker fit, including the sample means for the last two pre-parameters.

Next, we declare the model.      In this example there are no regressors      specified, and there is only one 
 latent process:
\begin{verbatim}
mdl <- NULL
mdl <- sigex.add(mdl    = mdl,
                 vrank  = seq(1,N),
                 class  = "varma",
                 order  = c(p.order,0),
                 bounds = NULL, 
                 name   = "process",
                 delta  = c(1,-1) )
# regressors:
mdl <- sigex.meaninit(mdl = mdl, data.ts = data.ts, d = 0)
\end{verbatim}
 In the second line  the main process is defined  through a call to \code{sigex.add}.  
 The list object \textsl{mdl} stores all our specifications about the model, but none of the parameter estimates.
 The second argument indicates the rank configuration for the trend;
    setting \textsl{vrank} equal to   \verb|seq(1,N)| ensures a full rank covariance matrix.
 The third argument  gives the model type as {\it varma};  
  the order $p=7$, $q=0$ of the VARMA is given in the fourth argument.  The fifth argument 
   provides  parameter bounds, and does not apply to this type of model, and hence is set to NULL.
     The sixth argument associates a label, and the last argument is for the scalar differencing operator
   $\delta^{(1)} (z)$, expressed as a vector of coefficients  \verb!c(1,-1)!.

   Now we move on to the next stage: model fitting.
 Here we demonstrate MLE using BFGS.  We begin by setting up parameter constraints; there are none in this case.
\begin{verbatim}
constraint <- NULL
psi.mle    <- psi.init
par.mle    <- sigex.psi2par(psi = psi.mle, mdl = mdl, data.ts = data.ts)
\end{verbatim}
The second line assigns \textsl{psi.mle} to the initial values of the pre-parameters -- this will speed up optimization.
 The third line converts this pre-parameter to the corresponding value of \textsl{param} via \code{sigex.psi2par}.
 Next, we run the MLE routine as follows:
\begin{verbatim}  
fit.mle <- sigex.mlefit(data.ts    = data.ts,
                        param      = par.mle,
                        constraint = constraint,
                        mdl        = mdl,
                        method     = "bfgs",
                        debug      = TRUE )
\end{verbatim}  
 With the debug option set to TRUE, the values of the divergence are printed to the console;  this is a way to track
  progress and understand what parts of the likelihood function are being searched.  Setting debug to FALSE turns off
  this screen output.    When the routine finally terminates (this can take a few minutes, depending on the machine),  we
   manage the output using
\begin{verbatim}
psi.mle <- sigex.eta2psi(eta = fit.mle[[1]]$par, constraint = constraint)
hess    <- fit.mle[[1]]$hessian
par.mle <- fit.mle[[2]]
\end{verbatim}
  The first line  ensures that if constraints have been used, the final $\eta$ is mapped to the corresponding $\psi$ via
   \code{sigex.eta2psi}.  In this case $\eta = \psi$ because there are no constraints.  We also extract the Hessian 
    matrix and the \textsl{param} corresponding to $\psi$ in the second and third lines.
    Next, we can check our model fit by examining the residuals.
\begin{verbatim}
resid.mle <- sigex.resid(psi = psi.mle, mdl = mdl, data.ts = data.ts)[[1]]
resid.mle <- sigex.load(data       = t(resid.mle),
                        start.date = start(data.ts),
                        period     = frequency(data.ts),
                        epithets   = colnames(data.ts),
                        plot       = TRUE )
resid.acf <- acf(x = resid.mle, lag.max = 4*period, plot = TRUE)$acf
\end{verbatim}

 The first line extracts the residuals corresponding to the MLE.    We want the first item in the list output of
  \code{sigex.resid}; the second item is an array of the fitted model's autocovariances.  
   The second call to \code{sigex.load} does formatting, and the last line computes sample autocorrelations of the residuals.  
 Next, we can examine    the condition numbers   via
\begin{verbatim}
log(sigex.conditions(data.ts = data.ts, psi = psi.mle, mdl = mdl))
\end{verbatim}
\begin{verbatim}
     [,1]       [,2]
eta2    0 -0.1349298
\end{verbatim}
The results indicate that the innovation covariance matrix is well-conditioned.  
 The model goodness-of-fit can be checked with
\begin{verbatim}
sigex.portmanteau(resids = resid.mle, lag = 4*period, nump = length(psi.mle))
\end{verbatim}
\begin{verbatim}
[1] 165.1150040   0.3534992
\end{verbatim}
 This code calls \code{sigex.portmanteau},  which yields the Portmanteau statistic for residual serial correlation
    discussed in  \cite{Lutkepohl2005};
     the second argument gives the number of lags to be used (48 in this case),
    and the third argument is the number of estimated parameters.
    The output indicates there is little  residual autocorrelation,  which is confirmed by inspection of the sample     autocorrelations.
   The call to \code{sigex.gausscheck} yields p-values for both series
    from the Shapiro-Wilks normality test; although there is some non-normality indicated, this does not invalidate the model.
\begin{verbatim}
sigex.gausscheck(resids = resid.mle)
\end{verbatim}
\begin{verbatim}
[1] 1.211720e-07 2.548542e-20
\end{verbatim}
We can examine the Hessian matrix,  and compute the t statistics for pre-parameters via

\begin{verbatim}
print(eigen(hess)$values)
tstats <- sigex.tstats(mdl,psi.mle,hess,constraint)
print(tstats)
\end{verbatim}

 If the Hessian is not positive definite (this can happen if nonlinear optimization terminates wrongfully at a saddlepoint),
  then the standard error is set to zero, and \code{sigex.tstats} returns $\pm \infty$ for each t statistic.
  In this case the Hessian is positive definite, and some of the pre-parameters are not significantly different from zero --
   we could refine the model by imposing zero constraints.
Finally, we bundle the analysis with the call
\begin{verbatim}
analysis.mle <- sigex.bundle(data.ts  = data.ts,
                            transform = "none",
                            mdl       = mdl,
                            psi       = psi.mle )
\end{verbatim}
 As an application,  we generate forecasts and aftcasts with uncertainty; also a midcast at time $t=1$ is generated
  for the New Orders series, since this value is missing.   Typically we unload the bundled model via
\begin{verbatim}
data.ts <- analysis.mle[[1]]
mdl     <- analysis.mle[[3]]
psi     <- analysis.mle[[4]]
param   <- sigex.psi2par(psi = psi, mdl = mdl, data.ts = data.ts)
\end{verbatim}
 This amounts to a relabeling.   Fifty forecasts and aftcasts are generated via
\begin{verbatim}
window <- 50
data.casts <- sigex.midcast(psi      = psi,
                            mdl      = mdl,
                            data.ts  = data.ts,
                            castspan = window )
extract.casts <- sigex.castextract(data.ts    = data.ts,
                                   data.casts = data.casts,
                                   mdl        = mdl,
                                   castspan   = window,
                                   param      = param )
\end{verbatim}
 The output of \code{sigex.midcast} is a list with two items:  the matrix of casts, and a 4-array of casting error
  covariances.   The $t$th column of the cast matrix contains a vector of casts for the $t$th index in the time
   series that has any missing values;  this takes into account the specified number of aftcasts and forecasts
   with the \code{castspan} variable.   Also note that these casts are for the series with all fixed effects removed. 
      This output can be formatted for use via  \code{sigex.castextract},  which inserts the casts in the correct
      places in the time series and adds an extended trend regression effect (it is extended forward and backward
      by the \code{castspan} value).  The output of \code{sigex.castextract} is a list with three items, in the
       same format as for \code{sigex.extract} output;  therefore the output can be used by
\code{sigex.graph}:
\begin{verbatim} 
dataPad.ts <- rbind(matrix(NA, nrow = window, ncol = N), 
                    data.ts, 
                    matrix(NA, nrow = window, ncol = N))
for(i in 1:N){
  plot(x    = ts(dataPad.ts[, i], start = c(1992, 1), frequency = 4),
       xlab = "Year", 
       ylab = "", 
       lwd  = 1, 
       col  = 1)
  sigex.graph(extract    = extract.casts,
              reg        = NULL,
              start.date = c(1992,1),
              period     = 4,
              series     = i,
              displace   = 0,
              castcol    = "black",
              fade       = 60 )
}
\end{verbatim}
 This code sets up grey shading for cast uncertainty  in \textsl{castcol}  and  \textsl{fade}.   
  For each component time series, the raw data is plotted  after first padding with NA
  before and after.    The first argument of \code{sigex.graph} takes the extracted list object \textsl{extract.casts};
  other fixed effects could be added in the second argument
  (these are typically not known at the times covered by the fore and aft windows, unless the regressors are
   simple to extrapolate).      The sixth
  argument allows for a vertical displacement in case one does not wish to overlay the extraction
  on top of the data -- setting to zero indicates no displacement.   The last argument determines the
  darkness of shading corresponding to the signal extraction uncertainty.
 This concludes the example.

\subsection{Petrol}

 We next examine  two monthly  petroleum time series: ``Industrial  Petroleum Consumption and OPEC Oil Imports,"
 or {\it petrol} for short.\footnote{Source: Department of Energy, Energy Information Administration.
   Both series cover the period   January 1973 through December 2016, and are
   seasonally adjusted (by DOE).    The first series,  ``Industrial Petroleum Consumption,"   is
   U.S. Product Supplied of Crude Oil and Petroleum Products, Monthly, in Thousand Barrels per Day.
	The second series, 	   ``OPEC Oil Imports,"   is   Petroleum Imports from OPEC, in  Thousand Barrels per Day. }
  We perform a bivariate analysis by fitting two simple models,   both of which are based on a trend-irregular 
   specification referred to as the Local Level Model (LLM) in 
   \cite{Harvey1989}.  Here there are no regressors 
    specified, and $K=2$, with $\{ S_t^{(1)} \}$ corresponding to a trend process and $\{ S_t^{(2)} \}$ a so-called
     irregular process, which is a  bivariate white noise.
        The trend is a bivariate random walk, i.e., $\delta^{(1)} (z) = (1 -z)$  with
      $\{ \underline{S}_t^{(1)} \}$ a white noise process.   
   These same series were analyzed with this model in
   \cite{mcelroy2020multivariatelinearpred}.

  	 After loading the data into a variable named \textsl{petrol},
  the   	 preliminary metadata is assigned as discussed in the prior example; a log transformation is used.
 We perform exploratory analysis by generating spectral density estimates;
 the function \code{sigex.specar}    generates vertical lines at frequencies that are multiples of
  $2 \pi$ divided by the \textsl{period} argument.
      The second argument of \code{sigex.specar} is a Boolean variable, indicating whether
       the data should be first differenced, i.e.,    apply $1-B$ to each series.
 Here we  set   \textsl{period} to $12$ in \code{sigex.specar} because  
 we are interested in the monthly frequencies:
\begin{verbatim}
par(mfrow = c(2, 1))
for(i in subseries){
    sigex.specar(data.ts   = data.ts,
                 diff      = FALSE,
                 subseries = i,
                 period    = period) 
}
\end{verbatim}
   The basic LLM is referred to as a ``related trends model,"  in contrast to a ``common trends model"
  considered subsequently, where it is enforced that $\Sigma^{(1)}$ has rank one.
  Consider the code
\begin{verbatim} 
mdl <- NULL
mdl <- sigex.add(mdl    = mdl,
                 vrank  = seq(1, N),
                 class  = "arma",
                 order  = c(0, 0),
                 bounds = NULL,
                 name   = "trend",
                 delta  = c(1, -1)
                )
mdl <- sigex.add(mdl    = mdl,
                 vrank  = seq(1, N),
                 class  = "arma",
                 order  = c(0, 0),
                 bounds = NULL,
                 name   = "irregular",
                 delta  = 1
                )
mdl <- sigex.meaninit(mdl = mdl, data.ts = data.ts, d = 0)
\end{verbatim}
  The  first call to  \code{sigex.add} defines the first latent process, labeled as ``trend,"   and the second such call gives the ``irregular" latent
   process.   Both latent processes have full rank.   Next, the  ``common trends model" is specified via
\begin{verbatim}
mdl2 <- NULL
mdl2 <- sigex.add(mdl    = mdl2,
                  vrank  = 1,
                  class  = "arma",
                  order  = c(0, 0),
                  bounds = NULL,
                  name   = "trend",
                  delta  = c(1,-1) )
mdl2 <- sigex.add(mdl    = mdl2,
                  vrank  = seq(1, N),
                  class  = "arma",
                  order  = c(0, 0),
                  bounds = NULL,
                  name   = "irregular",
                  delta  = 1 )
mdl2 <- sigex.meaninit(mdl = mdl2, data.ts = data.ts, d = 0)
\end{verbatim}
 where the second argument of \code{sigex.add} in the second line  
 has \textsl{vrank}  set to 1 instead of \verb!seq(1,N)!, as in \textsl{mdl}.
  
  Next, we fit the models,  beginning  with the related trends specification.
  The default parameter setting involves setting the pre-parameter vector to zero, via:
\begin{verbatim}
constraint <- NULL
par.mle <- sigex.default(mdl = mdl, data.ts = data.ts, constraint = constraint)
psi.mle <- sigex.par2psi(param = par.mle, mdl = mdl)
\end{verbatim}
The call to  \code{sigex.default} creates a \textsl{param} object \textsl{par.mle} corresponding to a pre-parameter 
  \textsl{psi.mle} of all zeroes -- there are 8  components in this case.   

  The collation and examination of results can be done in the manner described for the {\it ndc} example in Section \ref{sec:ndc}; for the related trends model,
  results are bundled into \textsl{analysis.mle}.
 In the case of the  common trends  model,
  we initialize at a given $\psi$ value  obtained from the first fitted model as follows:
   the pre-parameters associated with the trend innovation covariance matrix are altered
   to correspond to a full-correlation matrix (but with the same variances).   In this case
\[
  \Sigma = \left[ \begin{array}{cc}  \sigma_1^2  &  \rho \sigma_1 \sigma_2 \\ \rho \sigma_1 \sigma_2 & \sigma_2^2 
   \end{array} \right]   \mapsto
    \left[ \begin{array}{cc}  \sigma_1^2  &    \sigma_1 \sigma_2 \\   \sigma_1 \sigma_2 & \sigma_2^2 
   \end{array} \right], 
\]
  and this mapping is accomplished by setting the lower entry of $L$ in the GCD equal to $\sigma_2 / \sigma_1$,
    and $d_1$ is set equal to $\sigma_1^2$.  This is encoded with
\begin{verbatim}
cov.mat <- par.mle[[1]][[1]] %*% diag(exp(par.mle[[2]][[1]])) %*%
		 t(par.mle[[1]][[1]])
l.trend <- sqrt(cov.mat[2,2]/cov.mat[1,1])
d.trend <- cov.mat[1,1]
\end{verbatim}  
  Then the initialization to the common trends model is 
\begin{verbatim}
constraint <- NULL
psi.mle2 <- c(l.trend, log(d.trend), psi.mle[4:8])
par.mle2 <- sigex.psi2par(psi = psi.mle2, mdl = mdl2, data.ts = data.ts)
\end{verbatim}
   After fitting and managing output, we see that the common trends model is quite a bit worse
    (not surprising, given that the correlation $\rho$ is estimated at $.29$ in the related trends model),
   having a much higher divergence.  We can compare the fits via
\begin{verbatim}
test.glr <- sigex.glr(data.ts     = data.ts,
                      psi.nested  = psi.mle2,
                      psi.nesting = psi.mle,
                      mdl.nested  = mdl2,
                      mdl.nesting = mdl)
print(c(test.glr[1], 1-pchisq(test.glr[1], df = test.glr[2])))
\end{verbatim}
\begin{verbatim}
 [1] -21084.29      1.00
\end{verbatim}
  A likelihood comparison statistic 
  \citep{McElroy2016nonnested}
  for nested models is computed using \code{sigex.glr},
   and the p-value is reported in the second line.
   (However, this p-value is not correct  when the nested model
    arises at the boundary of the parameter space, as in this case.)
Finally, we consider trend extraction using either model.  For example, to load the second model we use
\begin{verbatim}
data.ts <- analysis.mle2[[1]]
mdl     <- analysis.mle2[[3]]
psi     <- analysis.mle2[[4]]
param   <- sigex.psi2par(psi = psi, mdl = mdl, data.ts = data.ts)
\end{verbatim}
  Here we use the matrix formula approach to signal extraction.   The first step is to get the matrix filters:
\begin{verbatim}
signal.trend <- sigex.signal(data.ts  = data.ts,
                             param    = param,
                             mdl      = mdl,
                             sigcomps = 1 )
signal.irr   <- sigex.signal(data.ts  = data.ts,
                             param    = param,
                             mdl      = mdl,
                             sigcomps = 2 )
\end{verbatim}
 There are two latent processes (trend and irregular), and the filters for these are obtained through the two calls to \code{sigex.signal};
  the last argument \textsl{sigcomps} of \code{sigex.signal} is a vector of indices denoting the composition of the desired signal.
  In this case,  \textsl{signal.trend} is composed of just the first latent process, so the value of \textsl{sigcomps} is $1$.
  The second step is to compute the extractions:
\begin{verbatim}
extract.trend <- sigex.extract(data.ts = data.ts,
                               filter  = signal.trend,
                               mdl     = mdl,
                               param   = param )
extract.irr   <- sigex.extract(data.ts = data.ts,
                               filter  = signal.irr,
                               mdl     = mdl,
                               param   = param )
\end{verbatim}
 In the first line, \textsl{extract.trend} is defined as a list object, with first item corresponding to a $T \times N$ matrix
  of trend extractions.  The second and third items of the list give upper and lower bounds of confidence intervals,
  based on $\pm 2$ square root signal extraction MSE.    Next, it is important to re-integrate fixed regression effects,
  and this is done via
\begin{verbatim}
reg.trend <- NULL
for(i in 1:N) {
  reg.trend <- cbind(reg.trend, sigex.fixed(data.ts = data.ts, 
                                            mdl     = mdl, 
                                            series  = i , 
                                            param   = param, 
                                            type    = "Trend" ) ) 
}
\end{verbatim}
  The function \code{sigex.fixed}  finds a specified regressor (named in the last argument) and multiplies by
   the corresponding parameter estimate,  for the component specified in the third \textsl{series}  argument.
   Here, the default linear trend line regressor is passed into the variable \textsl{reg.trend}.
   The results can be displayed using calls to \code{sigex.graph}, as in the following code:
\begin{verbatim}
trendcol <- "tomato"
fade     <- 40
par(mfrow=c(2,1),mar=c(5,4,4,5)+.1)
for (i in 1:N){
    plot(x    = data.ts[,i],
         xlab = "Year",
         ylab = colnames(data.ts)[i],
         ylim = c(min(data.ts[, i]), max(data.ts[, i])),
         lwd  = 2,
         yaxt = "n",
         xaxt = "n")
	sigex.graph(extract    = extract.trend,
	            reg        = reg.trend,
	            start.date = begin.date,
	            period     = period,
	            series     = i,
	            displace   = 0,
	            color      = trendcol,
	            fade       = fade)
	axis(1, cex.axis = 1)
	axis(2, cex.axis = 1)
}
\end{verbatim}
 This code sets up a red color for the trend in \textsl{trendcol}, and a shading percentage \textsl{fade}.
  For each component time series, the raw data is plotted (in logarithms) and overlaid with the trend;
  the first argument of \code{sigex.graph} takes the extracted list object \textsl{extract.trend},
  and any fixed effects that should be added must be included in the second argument.  The sixth
  argument allows for a vertical displacement in case one does not wish to overlay the extraction
  on top of the data -- setting to zero indicates no displacement.   The last argument determines the
  darkness of shading corresponding to the signal extraction uncertainty.
  
   Finally,  one can examine the   frequency response functions  for WK filters through \code{sigex.getfrf},
   which can also plot the real portions.
\begin{verbatim}
grid <- 1000
frf.trend <- sigex.getfrf(data.ts  = data.ts,
                          param    = param,
                          mdl      = mdl,
                          sigcomps = 1,
                          plotit   = TRUE,
                          grid     = grid )
\end{verbatim}
  The \textsl{grid} argument corresponds to a mesh of frequencies in $[0, \pi]$,  and the fourth argument of
   \code{sigex.getfrf} indicates the combination of components desired.   One can also examine the
   WK filter coefficients:
\begin{verbatim}   
len      <- 50
target   <- array(diag(N), c(N, N, 1))
wk.trend <- sigex.wk(data.ts  = data.ts,
                     param    = param,
                     mdl      = mdl,
                     sigcomps = 1,
                     target   = target,
                     plotit   = TRUE,
                     grid     = grid,
                     len      = len)
\end{verbatim}
     The \textsl{target} argument indicates that no linear combination of the trend is being considered here;
     \textsl{len} is set to $50$, so that the indices of the filter run from $-50 $ up to $50$.
     This finishes the example.

 \subsection{Housing Starts}

 Here we study ``New Residential Construction (1964-2012), Housing Units Started, Single Family Units," 
   or {\it  starts} for short,\footnote{The four series are   from
   the Survey of Construction of the U.S. Census Bureau, available at
     \url{http://www.census.gov/construction/nrc/how_the_data_are_collected/soc.html}.}
    corresponding to the four regions of  South, West, NorthEast, and MidWest.
    This example illustrates modeling a multivariate series with $8$ components, estimated
    with both MOM and MLE methods, and the extraction of signals with two methods (direct matrix and
     truncated WK).  We also show how functions of a signal, such as growth rates, can be obtained.

 The initial loading scripts are similar to the previous example.    Exploratory analysis indicates 
 we should consider a model with trend and seasonal effects;  here we consider a structural model
  with  $K=8$  latent processes:
\begin{verbatim}
mdl <- NULL
mdl <- sigex.add(mdl,seq(1,N),"arma",c(0,0),NULL,"trend",c(1,-2,1))
mdl <- sigex.add(mdl,seq(1,N),"arma",c(0,0),NULL,"first seasonal",c(1,-sqrt(3),1))
mdl <- sigex.add(mdl,seq(1,N),"arma",c(0,0),NULL,"second seasonal",c(1,-1,1))
mdl <- sigex.add(mdl,seq(1,N),"arma",c(0,0),NULL,"third seasonal",c(1,0,1))
mdl <- sigex.add(mdl,seq(1,N),"arma",c(0,0),NULL,"fourth seasonal",c(1,1,1))
mdl <- sigex.add(mdl,seq(1,N),"arma",c(0,0),NULL,"fifth seasonal",c(1,sqrt(3),1))
mdl <- sigex.add(mdl,seq(1,N),"arma",c(0,0),NULL,"sixth seasonal",c(1,1))
mdl <- sigex.add(mdl,seq(1,N),"arma",c(0,0),NULL,"irregular",1)
mdl <- sigex.meaninit(mdl,data.ts,0)
\end{verbatim}
Each of the eight latent processes is labeled, and has a distinct differencing polynomial,
 whose application produces a full-rank white noise; details of this
 model are discussed in  \cite{mcelroy2017multivariate}.     
 Next, we illustrate the MOM to obtain initial estimates.
\begin{verbatim}
mdl.mom     <- mdl
constraint  <- NULL
par.default <- sigex.default(mdl = mdl.mom, data.ts = data.ts, constraint = constraint)
par.mom     <- sigex.momfit(data.ts = data.ts, param = par.default, mdl = mdl.mom)
psi.mom     <- sigex.par2psi(param = par.mom, mdl = mdl.mom)
sigex.lik(psi     = psi.mom, 
          mdl     = mdl.mom, 
          data.ts = data.ts, 
          debug   = FALSE )
\end{verbatim}
\begin{verbatim}
[1] 6329.107
\end{verbatim}
 First, we define a new model object \textsl{mdl.mom}, and the subsequent initialization is standard in the second and third lines.
 The MOM estimates are obtained through \code{sigex.momfit} in the fourth line.   In the last call
 we evaluate the divergence,  obtaining $ 6329.107$.   
 
 Examination of the residuals indicates 
 that a substantial degree of serial dependence remains.   
   We can also examine the condition numbers, and observe that all the six seasonal latent processes seem to have
   reduced rank covariance matrices, indicating various types of seasonal co-integration.   
   We can modify the initial model to one that has
  reduced rank components, by using \code{sigex.reduce}:
\begin{verbatim}
thresh      <- -6.22
reduced.mom <- sigex.reduce(data.ts   = data.ts,
                            param     = par.mom,
                            mdl       = mdl.mom,
                            thresh    = thresh,
                            modelflag = TRUE )
mdl.mom     <- reduced.mom[[1]]
par.mom     <- reduced.mom[[2]]
psi.mom     <- sigex.par2psi(param = par.mom, mdl = mdl.mom)
sigex.lik(psi     = psi.mom,
          mdl     = mdl.mom,
          data.ts = data.ts,
          debug   = FALSE )
\end{verbatim}
\begin{verbatim}
[1] 6292.841
\end{verbatim}
  Here the threshold $\alpha = -6.22$, and the updated model is stored in \textsl{reduced.mom}, a list object.
  The first item in the list contains the updated information for the model -- in particular, the new rank configurations.
  These are entered into \textsl{mdl.mom} by extraction from 
   \textsl{reduced.mom}.  Then the  parameters are updated,
  and the last call just returns the divergence for the MOM estimates:  the value is $ 6292.841$.
Curiously, this is less than the value obtained previously for a nesting model;  however,  the previous divergence of
 $6329.107$  corresponds to a modification of the initial MOM fit where the covariance matrices have been rendered
 non-negative definite.  The reduced rank specification utilizes the same cholesky factors $L$, only with columns omitted
 corresponding to the rank configuration; hence it is possible to obtain a lower value of the divergence.
   
   We move to the MLE fitting at this point; initial values from the MOM fit are loaded through 
\begin{verbatim}
data.ts <- analysis.mom[[1]]
mdl     <- analysis.mom[[3]]
psi.mom <- analysis.mom[[4]]
par.mom <- sigex.psi2par(psi = psi.mom, mdl = mdl, data.ts = data.ts)
\end{verbatim}
 Fitting this reduced rank model can take a substantial amount of time, because 
 there are $67 $  pre-parameters to be estimated.   Optimization terminates at the divergence of
 $6152.607$, which is substantially better than that of the MOM estimate.
 Now when we examine the   condition numbers,  we see values of $-\infty$ (or very large negative values)
  corresponding to the complement of the rank configuration.

  The next stage of analysis is signal extraction, and both the matrix method and the WK method are illustrated
  in the script.  Here we focus on three latent processes: trend, seasonal, and seasonal adjustment.  Since the
  seasonal process is composed of the six atomic seasonals that occur at indices two through seven, 
  \textsl{sigcomps} equals \verb!seq(2,7)!.   Similarly, the setting is \verb!c(1,8)! for the seasonal adjustment.
  In the display we illustrate how the seasonal component can be plotted with a vertical displacement.
  In order to use \code{sigex.wkextract}, one sets the \textsl{grid} parameter to   determine the number of 
    mesh points in the interval $[-\pi,\pi]$ used as a Riemann approximation of the integral;
   the \textsl{horizon} parameter corresponds to $H$ (the number of desired aftcasts and forecasts),
    whereas  \textsl{target} is the array storing the coefficients of $\Xi (B)$.   Here  
  we set up the WK signal extraction via
\begin{verbatim}
grid    <- 7000	 
window  <- 50
horizon <- 0
target  <- array(diag(N), c(N, N, 1))
\end{verbatim}
 which specifies that the filter will be truncated to length $101$ (equal to twice \textsl{window} plus one)
 at the sample boundaries -- only $50$ forecasts and aftcasts will be generated.  Setting \textsl{horizon}
 to zero means that we will not forecast the signal ahead.  Finally, in the fourth line the \textsl{target}
  variable is set to the identity (as an array).   The calls to \code{sigex.wkextract} are then given by
\begin{verbatim}
extract.trend2 <- sigex.wkextract(psi      = psi,
                                  mdl      = mdl,
                                  data.ts  = data.ts, 
                                  sigcomps = 1,
                                  target   = target,
                                  grid     = grid,
                                  window   = window,
                                  horizon  = horizon,
                                  needMSE  = TRUE )
extract.seas2  <- sigex.wkextract(psi      = psi,
                                  mdl      = mdl,
                                  data.ts  = data.ts, 
                                  sigcomps = seq(2,7),
                                  target   = target,
                                  grid     = grid,
                                  window   = window,
                                  horizon  = horizon,
                                  needMSE  = TRUE )
extract.sa2    <- sigex.wkextract(psi      = psi,
                                  mdl      = mdl,
                                  data.ts  = data.ts, 
                                  sigcomps = c(1,8),
                                  target   = target,
                                  grid     = grid,
                                  window   = window,
                                  horizon  = horizon,
                                  needMSE  = TRUE )
\end{verbatim}
  The fourth argument in these function calls is \textsl{sigcomps}, and parallels the prior calls of \code{sigex.signal}.
  Note that whereas in the matrix approach we used two calls in succession (\code{sigex.signal} followed by \code{sigex.extract}),
  here there is a single call with output in the same list format as that provided by \code{sigex.extract}.
  The following plot in the script just compares the square root MSE for the trend components arising from both 
  signal extraction methods;  with \textsl{window} size of $50$ there is no discrepancy.   (The user can re-run the script
   with \textsl{window} set to $10$, to observe some discrepancy between the exact matrix results and the
    approximate WK results.)
 
 The final stage of this script is extraction of the trend growth rate, i.e., we extract $(1-B) S_t^{(1)}$.   Note
 that this is an I(2) trend,  so the growth rate is a random walk -- however,  the dynamics of an extracted component
 in general do not match those  of  the specified process  (see discussion in 
 \cite{McElroy2012}).
 The first code block extracts the trend growth rate and MSE directly from the output of \code{sigex.signal}
 stored in \textsl{signal.trend}, and we do not cover this in detail, instead focusing on the second code block
 which produces approximately the same results via the WK filter.   Noting that the growth rate polynomial
  \textsl{gr.poly} has already been set to \code{c(1,-1)},  we proceed to
\begin{verbatim}
grid     <- 70000  # need large grid value to get accuracy
window   <- 100
horizon  <- 0
gr.array <- array(t(gr.poly) \%x\% diag(N), c(N, N, p+1))
reg.gr   <- array(0, c(T, N))
for (k in 1:N) { 
    reg.gr[, k] <- filter(x     = reg.trend[, k],
                         filter = gr.poly,
                         method = "convolution",
                         sides  = 1) 
}
\end{verbatim}
In the first line we take a large \textsl{grid} value to get decent accuracy, and take a larger truncation \textsl{window}
 of $100$.  In the fourth line,  the target is defined by \textsl{gr.array} by writing \textsl{gr.poly} in an array.
 Finally, we  need to modify the trend regressors so that they correctly correspond to the trend growth rate;
 this is accomplished by filtering the polynomial trend with \textsl{gr.poly}.
 (Recall that \textsl{reg.trend} was defined above, and already incorporates the regression parameters.)
 Next,  we call
\begin{verbatim}
extract.trendgr2 <- sigex.wkextract(psi      = psi,
                                    mdl      = mdl,
                                    data.ts  = data.ts,
                                    sigcomps = 1,
                                    target   = gr.array,
                                    grid     = grid,
                                    window   = window,
                                    horizon  = horizon,
                                    needMSE  = TRUE )
\end{verbatim}
 which can take some time because of the high value of \textsl{grid}.  Note that \textsl{gr.array} is now in the
  \textsl{target} argument.   The output is plotted in the usual way, displaying the trend growth rate for the four
   series, with appropriate uncertainty.    The script repeats the exercise for the seasonal adjustment,
    and the output can be compared to the direct matrix output.   This concludes the example.

\subsection{Business Formation Statistics}


 Next, we study a weekly univariate series: the business applications   component of   the
  ``Business Formation Statistics,"\footnote{Weekly data as of week 27  of 2020, downloaded on  
   July 14, 2020 (11:27 AM), U.S. Census Bureau, obtained from 
 \url{https://www.census.gov/econ/bfs/index.html}.} or \textit{bfs} for short.
  We consider the non-seasonally adjusted data at the national level.
   This  component is the main business applications series, which captures the weekly flow of
    IRS applications for Employer Identification Numbers, mainly for business purposes.
  For the metadata specifications we enter
\begin{verbatim}
begin  <- c(2006, 1)
end    <- c(2020, 27)
period <- 52
\end{verbatim}
  which uses the fallacious \textsl{period} of $52$.  The actual period should be approximately $365.25/7 \approx
   52.17857$, but   \textsl{Ecce Signum} requires integer values of \textsl{period}.
   This is not entirely satisfactory, because a $53$rd week is present every five to seven years; 
   we present some ways below to circumvent this difficulty.
   
   In loading the series, we select one of four national series corresponding to business applications, and choose
    a log transformation.   In order to specify the holiday regressors, we need to know the calendar date
    (in  month-day-year format) for the beginning and end of the series, i.e., the first day of the first week
    and the last day of the last week.   The function \code{weekly2date} provides this:
\begin{verbatim}
first.day  <- 1
all.date   <- weekly2date(first.day = first.day, begin = begin, T = T)
start.date <- all.date[[1]]
end.date   <- all.date[[2]]
\end{verbatim}
 Weekly data typically corresponds to measurements over a seven-day period beginning with Sunday, in which 
 case \textsl{first.day} is set equal to $1$,  but non-standard week configurations can be accommodated as well.
 Having extracted the output in lines three and four,  we can load the holiday regressors.
One supplies an ASCII file  whose every row contains a calendar date for the relevant holiday, covering a broad range of years.
 (Because we later  remove a long-term mean of the holiday regressor to avoid confounding with seasonal
  effects,  we recommend providing these dates for several centuries.)  
   These date files can be constructed from the Internet, or the \code{datefinder} program can be used.
   For Easter we utilize the following code; note \code{easter500}\footnote{This is a conversion of the U.S.Census Bureau's list of Easter  dates for a 500 year period (1600-2099): \texttt{https://www.census.gov/srd/www/genhol/easter500.html}} comes preloaded with the package.
\begin{verbatim}
easter.reg <- gethol(hol.dates  = easter500, 
                     hol.fore   = 7, 
                     hol.aft    = 0, 
                     start.date = start.date, 
                     end.date   = end.date )
\end{verbatim}
 Regressors can simply be created as a daily
 time series by \code{gethol}, and then later converted to a weekly regressor.  Here we consider a window
 of $7$ days before and $0$ days after, so that the holiday effect is presumed to constitute eight days, including
 the entire prior week.   Note that in the call to \code{gethol},  both \textsl{start.date} and \textsl{end.date}
 are required.
 
 For \textit{bfs} we consider several holiday effects: all the ten federal holidays as well as
  Easter and Black Friday.  In each case we take the actual date of the holiday rather than
  its observed date (the observation dates are more difficult to determine, but this could be done).
       Note that holidays that do not have moving dates (such as Independence Day,
   Veteran's Day, and Christmas)  constitute a deterministic annual seasonal effect;
   because \code{gethol} removes a long-term mean over many years (so as to avoid confounding
   with seasonal effects), such regressors become zero.   (We still include these three holidays
   in our analysis to demonstrate below that they are automatically removed from the model.)
  
   Next, the daily regressors are converted to weekly flow regressors by first embedding
   with \code{sigex.daily2weekly} followed by averaging over each week:
\begin{verbatim}
easter.reg <- sigex.daily2weekly(data.ts    = easter.reg,
                                 first.day  = first.day,
                                 start.date = start.date)
easter.reg <- rowSums(easter.reg)/7
\end{verbatim}
 The averaging in the third line corresponds to the construction of monthly and quarterly regressors in
  X-12-ARIMA  described in \cite{FindleyMonsellBellOttoChen1998},
    whereby the proportion of the effect (over the entire window) that falls in any given week is declared
  to be the value of the regressor for that week.
   
   After these specifications,  we fit a baseline model that involves no fixed effects,  but models the 
   serial dependence with a SARMA specification with order one for the AR, MA, seasonal AR, and
    seasonal MA polynomials:
\begin{verbatim}
mdl <- NULL
mdl <- sigex.add(mdl    = mdl,
                 vrank  = seq(1,N),
                 class  = "sarma",
                 order  = c(1,1,1,1,52),
                 bounds = NULL,
                 name   = "process",
                 delta  = 1)
mdl <- sigex.meaninit(mdl = mdl, data.ts = data.ts, d = 0)
\end{verbatim}
 In particular, we do not specify any differencing even though the level of the series seems to change --
  gradual (but transitory) level changes are not inconsistent with a stationary specification.
      The divergence is $-2076.881$,  and the fit seems reasonably good.
    Our next model incorporates the  the ten federal holidays along with Easter and Black Friday,
    although Independence Day, Veteran's Day, and Christmas are not included by the call to
    \code{sigex.reg}  (which checks for regressors that are numerically zero after differencing).
    For instance,
\begin{verbatim}
mdl <- sigex.reg(mdl = mdl, 
                 series = 1, 
                 reg = ts(data      = as.matrix(xmas.reg), 
                          start     = start(xmas.reg), 
                          frequency = period, 
                          names     = "Xmas") )
mdl <- sigex.reg(mdl = mdl, 
                 series = 1, 
                 reg = ts(data      = as.matrix(black.reg), 
                          start     = start(black.reg), 
                          frequency = period, 
                          names     = "BlackFriday") )
\end{verbatim}
 will make no modification to \textsl{mdl} in the first call, whereas the second call adds the Black Friday regressor.
   So this specification has seven federal holidays, plus two other holidays and the trend constant, for a total
   of $10$ regressors.
      This improved model is fitted, producing a divergence of $ -2089.925$,
       but the routine terminates at a saddlepoint    (the Hessian has a negative eigenvalue).
 We refine the model by restricting   ourselves to the New Year's (NY) and
    Martin Luther King (MLK) holidays.   
   At this point, we observe in the fitted residuals that there is a substantial outlier in the first week of 2012,
   which is not explained by the New Year's regressor.  We can treat this as an additive outlier, 
    which is handled via
\begin{verbatim}
AO.times  <- 314
dataNA.ts <- data.ts
dataNA.ts[AO.times] <- NA
\end{verbatim}
  This defines a new time series object \textsl{dataNA.ts} that is identical to \textsl{data.ts} but has NA inserted
  wherever an additive outlier occurs.  Our third refined model is therefore
\begin{verbatim}
mdl <- NULL
mdl <- sigex.add(mdl = mdl,
                 vrank = seq(1,N),
                 class = "sarma",
                 order = c(1, 1, 1, 1, 52),
                 bounds = NULL,
                 name = "process",
                 delta = 1)
mdl <- sigex.meaninit(mdl = mdl, data.ts = dataNA.ts, d= 0)
mdl <- sigex.reg(mdl = mdl, 
                 series = 1, 
                 reg = ts(data      = as.matrix(nyd.reg), 
                          start     = start(nyd.reg), 
                          frequency = period, 
                          names     = "NewYearDay") )
mdl <- sigex.reg(mdl = mdl, 
                 series = 1, 
                 reg = ts(data      = as.matrix(mlk.reg), 
                          start     = start(mlk.reg), 
                          frequency = period, 
                          names     = "MLK") )
\end{verbatim}
     This model is fitted     and a lower divergence ($-2329.284$)   is obtained -- 
     this is theoretically impossible (because the model is nested),
    but occurs due to wrongful termination of the optimization search in the larger model.
    This is not an uncommon situation in our experience, when working with time series models with
    many parameters, and seems worth mentioning.   For this final model, we get a positive definite Hessian
    and hence t statistics for the parameters, confirming the significance of all the variables.
    Residual analysis indicates the model seems to be a good fit.
    We record the midcast for time $t=314$ via
\begin{verbatim}
data.casts <- sigex.midcast(psi      = psi.mle,
                            mdl      = mdl,
                            data.ts  = dataNA.ts,
                            castspan = 0 )
\end{verbatim}
     and bundle the analysis using \textsl{dataNA.ts}.   The final stage of analysis is to filter the data.
     The relevant fixed effects are loaded into variables via
\begin{verbatim}
reg.trend  <- sigex.fixed(data.ts = data.ts,
                          mdl     = mdl,
                          series  = 1,
                          param   = param,
                          type    = "Trend" )
reg.nyd    <- sigex.fixed(data.ts = data.ts,
                          mdl     = mdl,
                          series  = 1,
                          param   = param,
                          type    = "NewYearDay" )
reg.mlk    <- sigex.fixed(data.ts = data.ts,
                          mdl     = mdl,
                          series  = 1,
                          param   = param,
                          type    = "MLK" )
dataLIN.ts <- data.ts - ts(data      = reg.trend + reg.nyd + reg.mlk, 
                           start     = start(data.ts), 
                           frequency = period)
\end{verbatim}
 The last instruction creates \textsl{dataLIN.ts},  consisting of the original time series with the fixed effects
  removed, i.e., $\widehat{Y}_t$.   This de-meaned series  shall be adjusted for the AO, and then filtered.
  We utilize a set of nonparametric filters to extract trend, seasonal, and non-seasonal (or seasonally adjusted)
  latent processes, implemented in \code{x11filters}.    Details of these filters are given in Appendix B. 
\begin{verbatim}  
week.period  <- 365.25 / 7
half.len     <- floor(week.period / 2)
x11.filters  <- x11filters(period = week.period, p.seas = 1)
trend.filter <- x11.filters[[1]]
seas.filter  <- x11.filters[[2]]
sa.filter    <- x11.filters[[3]]
shift        <- (dim(sa.filter)[3] - 1) / 2
\end{verbatim}
     The first line defines the non-integer period, which \code{x11filters} is designed to accommodate.
     The call in the third line returns a three-item list, containing the three filters, each of which is
      a symmetric sequence of numbers.   Next, we call \code{sigex.adhocextract}:
\begin{verbatim} 
trend.comp <- sigex.adhocextract(psi     = psi,
                                 mdl     = mdl,
                                 data.ts = dataNA.ts,
                                 adhoc   = trend.filter,
                                 shift   = half.len,
                                 horizon = 0,
                                 needMSE = TRUE )
sa.comp    <- sigex.adhocextract(psi     = psi,
                                 mdl     = mdl,
                                 data.ts = dataNA.ts,
                                 adhoc   = sa.filter,
                                 shift   = shift,
                                 horizon = 0,
                                 needMSE = TRUE )
AO.errs <- dataLIN.ts[AO.times] - data.casts[[1]]
sa.comp[[1]][AO.times] <- sa.comp[[1]][AO.times] + AO.errs
\end{verbatim}
 In the second call, note that we apply the method to \textsl{dataNA.ts}, which was defined previously
 as the de-meaned series with an NA inserted for the AO.   The method will automatically generate the
 appropriate cast for the NA, filter the resulting casted series,  and return the output along with 
 appropriate uncertainty.   The AC filter \textsl{trend.filter} is inputted, along with the \textsl{shift} parameter
 set to the middle position of the filter vector -- this corresponds to a symmetric filter.  
 The penultimate line computes the casting error $\varepsilon_t$ at the AO time $t = 314$,  and this is added
 onto the seasonal adjustment in the last line -- because an AO effect belongs with the irregular in the
 seasonal adjustment extraction.
 
 The final parts of the example display the extractions.   The illustration is complete.

\subsection{New Zealand Arrivals}
    

  Our last  illustration considers  a daily immigration series  (\textit{nz} for short):  
  New Zealand residents arriving in New Zealand after an absence of less than 12 months,      covering the period 
  January 1, 2008     through July 31, 2012.\footnote{ The series consists of New Zealand 
  residents arriving in New Zealand after an absence of less than 12 months. 
  These are public use data  produced by Stats New Zealand via a customized extract, and correspond to a portion of the
  ``daily border crossings - arrivals" tab (Total) of the Travel category in the  Covid-19 portal:  
  \url{https://www.stats.govt.nz/experimental/covid-19-data-portal}.}   The most obvious dynamics are
       the annual movements due to seasonality;  more subtle are the weekly dynamics.        
      Related time series (of longer sample size) were investigated in 
      \cite{McElroyJach2019};
 here  we find that the sample autocorrelation functions for the   series is somewhat different
      for each of the seven weekly series corresponding to each day-of-the-week.  
      This phenomenon indicates that the daily series -- analogously to the daily retail series analyzed in \cite{McElroyCasting}  --
      can be modeled by a vector  embedding as a $7$-variate weekly series.
 Some initial processing of the daily data is needed:
\begin{verbatim}
n.months <- dim(imm)[1]/32
imm <- imm[-seq(1,n.months)*32,]    # strip out every 32nd row (totals)
imm <- matrix(imm[imm != 0],ncol=6) # strip out 31st False days
\end{verbatim}
 The original data is grouped into ``months" of 31 days, with extra days of value zero added for months of shorter length;
  our code removes these features, as well as a monthly summation.  
  Next, a set of customized regressors, \textsl{NZregs} comes with package.
 The first  column is  a regressor that captures  a moving holiday effect for Easter, and the next six columns
  correspond to school holidays for public schools  in New Zealand:  an effect is estimated 
   for the start and end of the holiday for the
first three school vacations in each year.\footnote{The school holiday regressor for the first day of
the vacation period is set to one for that day and one for the day after, zero otherwise. 
 The school holiday regressor for the
last day of the vacation period is set to one for that day and one for the day before, zero
otherwise.  The school holiday regressors are centered by removing the daily mean of each of the regressors over
the entire series.} The last three columns correspond to Y2K effects, and will not be used.  For metadata we enter
\begin{verbatim}
start.date <- c(9, 1, 1997)
end.date   <- day2date(day = dim(imm)[1] - 1, start.date = start.date)
period     <- 365
\end{verbatim}
 The full data begins on September 1, 1997 and ends July 31, 2012, though we will be interested in a restricted range.
 Following are some calendrical calculations, and a call to \code{sigex.load}:
\begin{verbatim}
dataALL.ts <- sigex.load(data       = imm,
                         start.date = begin,
                         period     = period,
                         epithets   = c("NZArr", "NZDep", "VisArr",
                                        "VisDep", "PLTArr", "PLTDep"),
                         plot       = TRUE )
\end{verbatim}
There are six time series in the data set, though we will be focused on just the first one ``NZArr," described above.
 We restrict the data span and focus on the first series via
 \begin{verbatim}
transform  <- "log"
aggregate  <- FALSE
subseries  <- 1
range      <- list(c(2008, 1), end)
dataONE.ts <- sigex.prep(data.ts   = dataALL.ts,
                         transform = transform,
                         aggregate = aggregate,
                         subseries = subseries,
                         range     = range,
                         plot      = TRUE )
\end{verbatim}
 We can examine spectral plots to get an idea of the time series dynamics.  
 Also we set the first day to be Sunday, and embed as a $7$-variate weekly time series
 via a call to \code{sigex.daily2weekly}:
 \begin{verbatim}
first.day <- 1
data.ts   <- sigex.daily2weekly(data.ts = dataONE.ts, 
                                first.day = first.day, 
                                start.date = start.date)
plot(data.ts)
\end{verbatim}
  As a result $N= 7$ and $T = 240$,  which yields an initial NA (because the first data value is on Monday, but
   the week begins on Sunday).  Also the data ends on a Sunday, so the  last week has six NA values.
   This is a classic example of a ragged edge data pattern.
   Next, we  define appropriate subspans of the holiday regressors via
 \begin{verbatim} 
s <- (range[[1]][1] - begin[1]) * period + (range[[1]][2] - begin[2]) + 1
e <- (range[[2]][1] - begin[1]) * period + (range[[2]][2] - begin[2]) + 1
times <- s:e 
easter.reg   <- NZregs[times,1]
school1.reg  <- NZregs[times,2]
school1e.reg <- NZregs[times,3]
school2.reg  <- NZregs[times,4]
school2e.reg <- NZregs[times,5]
school3.reg  <- NZregs[times,6]
school3e.reg <- NZregs[times,7]
\end{verbatim}
 We have labeled the regressors corresponding to Easter and the three school holiday effects (both the beginning and end of each holiday period).   These are the regressors for daily data, and need to be embedded as weekly series:
\begin{verbatim} 
easter.reg <- sigex.daily2weekly(data.ts    = easter.reg, 
                                 first.day  = first.day, 
                                 start.date = start.date )
easter.reg[is.na(easter.reg)] <- 0
\end{verbatim}
The last line replaces any  ragged edge NA values (arising from embedding) with zero, which is appropriate because the
 corresponding data values are also missing.  
 
 Now we come to the model specification.  The data has a slight upwards trend drift, and there is strong annual correlation
  present.  Although a stationary model can be entertained, we found a fairly good model fit given by a SVARMA for
  data differenced to stationarity by applying annual differencing $1 - L^{52}$.  (Recall $L$ is the weekly lag operator,
   and there are approximately 52 weeks per year.)   After trying different possibilities, we consider nonseasonal $q=1$ and
    seasonal $P = 1$, which is specified by the code
 \begin{verbatim} 
mdl <- NULL
mdl <- sigex.add(mdl    = mdl,
                 vrank  = seq(1,N),
                 class  = "svarma",
                 order  = c(0, 1, 1, 0, 52),
                 bounds = NULL,
                 name   = "process",
                 delta  = c(1, rep(0, 51), -1) )
mdl <- sigex.meaninit(mdl = mdl, data.ts = data.ts, d = 0)
\end{verbatim}
 The specification of the regressors takes more coding, and we show the code for Easter (the index $i$ loops over 
  the days of the week):
 \begin{verbatim}
 mdl <- sigex.reg(mdl    = mdl, 
                  series = i, 
                  reg    = ts(data      = as.matrix(easter.reg[,i]),
                              start     = start(easter.reg),
                              frequency = frequency(easter.reg),
                              names     = "Easter-day") )
\end{verbatim}
 Next we consider model fitting.  Here we make use of parameter constraints for the regressors: because
  the same  holiday effect is present for each day of the week (recall that the weekly regressors are
   derived from a single daily regressor), the parameter estimates should be constrained to be identical.
We use the \code{sigex.constrainreg} function to do this:  the \textsl{regindex} argument specifies
   indices $J_1, J_2, \ldots, J_N$ that delineate which regressors are being considered for each series.
     For instance, $J_1 = \{ 1, 3, 4 \}$
  would correspond to the trend, school1, and school1e regressors for the first (Sunday) series.
  The \textsl{combos} argument provides the linear combinations of these regressors, but when
   set to NULL instead it is enforced that all the regression parameters are the same.
\begin{verbatim}   
constraint <- rbind(constraint,
    sigex.constrainreg(mdl,data.ts,list(2,2,2,2,2,2,2),NULL))
constraint <- rbind(constraint,
    sigex.constrainreg(mdl,data.ts,list(3,3,3,3,3,3,3),NULL))
constraint <- rbind(constraint,
    sigex.constrainreg(mdl,data.ts,list(4,4,4,4,4,4,4),NULL))
constraint <- rbind(constraint,
    sigex.constrainreg(mdl,data.ts,list(5,5,5,5,5,5,5),NULL))
constraint <- rbind(constraint,
    sigex.constrainreg(mdl,data.ts,list(6,6,6,6,6,6,6),NULL))
constraint <- rbind(constraint,
    sigex.constrainreg(mdl,data.ts,list(7,7,7,7,7,7,7),NULL))
\end{verbatim}
 So the first call here identifies in \textsl{regindex} the Easter regressor for each of the seven series,
  and \textsl{combos} set to  NULL indicates that these parameters will be equal.
  The next call does the same for the school1 regressor, and so forth; the first regressor is the trend,
   and this  remains unconstrained.   The output of \code{sigex.constrainreg} is a \textsl{constraint} matrix,
    which can be then passed into the MLE fitting routine.  In this case the nonlinear optimization
    can take a long time -- several days -- and we just report the output in the script.
    The remaining code lines review diagnostics, as discussed in previous illustrations.
    
    Moving on to signal extraction, we consider applying a daily seasonal adjustment filter that 
    we embed as a $7$-variate weekly filter.  Consider the code
\begin{verbatim}    
sa.hifilter <- c(1, rep(2, 365), 1) / (2 * 365)
len      <- 183
hi.freq  <- 7
low.freq <- 1
shift.hi <- len
out      <- sigex.hi2low(filter.hi = sa.hifilter,
                         hi.freq   = hi.freq,
                         low.freq  = low.freq,
                         shift.hi  = shift.hi )
sa.lowfilter <- out[[1]]
shift.low    <- out[[2]]
\end{verbatim}
The daily seasonal adjustment filter \textsl{sa.hifilter} is an extension of the classic crude trend filter
 (known as the $2 \times 12$ filter)  of X11  (discussed in  \cite{McElroyPolitis2020}) from monthly to daily frequency;
  it should suppress all daily frequencies of the form $ 2\pi j/365$ for $1 \leq j \leq 183$, while also
   linear trends.   This is embedded with a call to \code{sigex.hi2low}, where the high and low  frequencies
    are indicated and the \textsl{shift.hi} parameter is $ c = 183$, corresponding to a symmetric filter.
    The output is the $7$-variate weekly filter \textsl{sa.lowfilter}, with shift parameter $C = 27$ given by \textsl{shift.low}.

    Next, this filter is applied using a standard call of \code{sigex.adhocextract}.  
    Recall that this automatically supplies missing value imputations for the ragged edges, and does forecast and aftcast
     extension for application of the filter.      However, this should be transformed back to a daily series,
     so that we can view the daily series with its seasonal adjustment:
 \begin{verbatim} 
daily.ts <- ts(data      = sa.low[[1]],
               start     = start(dataONE.ts),
               frequency = frequency(dataONE.ts) )
sa.hi.daily[[1]] <- sigex.weekly2daily(data.ts   = daily.ts,
                                       first.dat = first.day )
\end{verbatim}
     The function \code{sigex.weekly2daily} transforms the series, using the indicated value of \textsl{first.day}.
 We repeat this for the lower and upper values of the confidence intervals (not shown).  This can now be displayed,
 but we first generate another signal for comparison -- we remove only the weekly dynamics.  This is done
 by considering a simple filter annihilating weekly effects from the daily data, namely $(1 + B + \ldots + B^6)/7$,
  which we call the Trading Day (TD) filter.  The relevant code is
 \begin{verbatim}
td.hifilter <- rep(1, 7) / 7
len      <- 3
hi.freq  <- 7
low.freq <- 1
shift.hi <- len
out <- sigex.hi2low(filter.hi = td.hifilter,
                    hi.freq   = hi.freq,
                    low.freq  = low.freq,
                    shift.hi  = shift.hi )
td.lowfilter <- out[[1]]
shift.low    <- out[[2]]
\end{verbatim}
 Now $c = 3$ and $C = 1$ for the TD filter.   Before plotting, the trend must be computed.  A subtlety is that
  we have $7$ trends, one for each day of week, which  therefore combine  long-term trend effects and 
   day-of-week effects.   The mean of the $7$ parameter values corresponds to the overall mean of each week,
   and the residual is the day-of-week effect in excess of the mean.  So   we define the daily trend to be the output of
   applying \textsl{td.hifilter} to the regression mean, i.e.,
\begin{verbatim}
reg.td <- NULL
for (i in 1:N){
  reg.td <- cbind(reg.td, sigex.fixed(data.ts = data.ts,
                                      mdl     = mdl,
                                      series  = i,
                                      param   = param,
                                      type    = "Trend") )
}
reg.td <- rbind(rep(0, N), reg.td)
reg.td <- ts(data      = sigex.weekly2daily(data.ts = reg.td, first.day = first.day),
             start     = start(dataONE.ts),
             frequency = period)
reg.trend <- filter(x      = reg.td,
                    filter = td.hifilter,
                    method = "convolution",
                    sides  = 1)
reg.trend <- as.matrix(reg.trend[8:length(reg.td)])
\end{verbatim}
 First \textsl{reg.td} is computed, which is the trend means for each day-of-week expressed as a daily time series.
  We augment by one prior week's values, which would be all zero, and then convert to a daily series.
  Next, \textsl{reg.trend} is obtained by filtering with $(1 + B + \ldots + B^6)/7$, and finally remove the NA values
   (resulting from using \textsl{filter}, which does not extend the input series).  The illustration is completed
   with a display:  the seasonal adjustment contains some mild weekly effects (because the filter is designed to 
    remove daily, and not weekly frequencies), whereas the TD-adjusted component has annual movements but no
     weekly effects.

\section*{Acknowledgments}

This report is released to inform interested parties of research and to encourage discussion.  The views expressed on
statistical issues are those of the authors and not  those of the U.S. Census Bureau.

\bibliographystyle{apalike}
\bibliography{bibliography}

\appendix


\section{Definition of Cyclical Processes}
\label{app:cyclical-process}
 
\subsection{Butterworth Cycle}

The Butterworth cyclical process is described in 
\cite{HarveyTrimbur2003},
   and in the first order case is an ARMA(2,1) process whose
 parameters are governed by a persistence $\rho$ and a frequency $\omega$. 
   The order $n$ {\em Butterworth cyclical process} is defined via
\[
  {(1 - 2 \, \rho \, \cos (\omega) B + \rho^2 \, B^2)}^n \, x_t = { ( 1 - \rho \, \cos (\omega) B) }^n \, \epsilon_t,
\]
 for $\epsilon_t \sim \mbox{WN} (0, \sigma^2)$.  It is known that the causal moving average representation corresponds
 to $x_t = {\Psi (B) }^n \, \epsilon_t$ with  
\[
  \Psi (B) = \sum_{j \geq 0 } \psi_j \, B^j = \frac{ 1 - \rho \, \cos (\omega) \, B }{  1 - 2 \, \rho \, \cos (\omega ) \, B + \rho^2 \, B^2 }
 \qquad  \mbox{and} \qquad \psi_j = \rho^j \,  \cos (\omega j ).
\]
 Clearly, as $\rho$ increases towards one the cyclical dynamics become increasingly regular,
   and the impulse response function when $n=1$
 becomes that of perfect sinusoid.  Moreover, the AR(2) operator matches the case of an atomic process with complex conjugate root
 $e^{ \pm i \omega }$, as $\rho \tends 1$.  The period of the cyclical effect is approximately $2 \pi / \omega$.

 Stabilization is determined by minimizing $f(\lambda) = { | \Psi (e^{-i \lambda} ) |}^{2n}$.  Taking the logarithm and differentiating yields
\[
  \frac{ - 2 n \, \rho \, \sin (\omega + \lambda) }{ 1 + \rho^2 - 2 \, \rho \, \cos (\omega + \lambda) }
  - \frac{ - 2 n \, \rho \, \sin (\omega - \lambda) }{ 1 + \rho^2 - 2 \, \rho \, \cos (\omega -  \lambda) }
  + \frac{ 2 n \, \rho \, \cos (\omega) \, \sin (\lambda) }{ 1 - 2 \, \rho \, \cos (\omega) \, \cos (\lambda) + \rho^2 \, { \cos (\omega) }^2 }.
\]
Setting this equal to zero and simplifying, we must solve
\begin{align*}
 0 & = (2 \, \rho \, \cos (\omega) \, \sin (\lambda) ) \, \left( 1 + 4 \, \rho^2 \, { \cos (\omega)}^2 + \rho^4 - 4 \, \rho \, (1 + \rho^2) \,
	\cos (\omega) \, \cos (\lambda) + 2 \, \rho^2 \, \cos (2 \lambda) \right)  \\
	& \quad  - \left(1 + \rho^2 \, { \cos (\omega)}^2 - 2 \, \rho \, \cos (\omega) \, \cos (\lambda) \right) \,
   	\left( 4 \, \rho \, (1 + \rho^2) \, \cos (\omega) \, \sin (\lambda) - 4 \, \rho^2 \, \sin (2 \lambda) \right) \\
& =  2 \, \rho \, \cos (\omega)  \, \left( 1 + 4 \, \rho^2 \, { \cos (\omega)}^2 + \rho^4 \right) \, \sin (\lambda)
 	+ 4 \, \rho^3 \, \cos (\omega) \, \sin (\lambda) \, \cos (2 \lambda)   \\
	& \quad - 4 \, \rho \, (1 + \rho^2 \, { \cos (\omega) }^2  ) \, (1 + \rho^2) \, \cos (\omega) \, \sin (\lambda)
	+ 8 \, \rho^2 \, (1 + \rho^2 \, { \cos (\omega) }^2 ) \, \sin (\lambda) \, \cos (\lambda)  \\
	& \quad - 16 \, \rho^3 \, \cos (\omega) \, \sin (\lambda) \, {\cos (\lambda) }^2.
\end{align*}
Hence $\lambda = 0, \pi$ are critical points; but if $\lambda \in (0, \pi)$ we can divide out by $\sin (\lambda)$, obtaining
\begin{align*}
  0 & = a_0 + a_1 \, z + a_2 \, z^2 \\
 a_0 & =  2 \, \rho \, \cos (\omega)  \, \left( 1 + 4 \, \rho^2 \, { \cos (\omega)}^2 + \rho^4 \right) 
	- 4 \, \rho^3 \, \cos (\omega) 
 	 - 4 \, \rho \, (1 + \rho^2 \, { \cos (\omega) }^2  ) \, (1 + \rho^2) \, \cos (\omega)  \\
	& = 2 \, \rho \, \cos (\omega)  \, \left( 2 \, \rho^2 \, { \cos (\omega)}^2 \, (1 - \rho^2) - 1 - 4 \, \rho^2 \right) \\
 a_1 & = 8 \, \rho^2 \, (1 + \rho^2 \, { \cos (\omega)}^2 ) \\
 a_2 & = -8 \, \rho^3 \, \cos (\omega)
\end{align*}
 with $z = \cos (\omega)$.   We apply the quadratic formula: the discriminant simplifies to 
\[
{\sin (\omega)}^2 \,	\left( { \sin (\omega) }^2 + {(1 - \rho^2)}^2 \, { \cos (\omega) }^2 \right),
\]
  and the cosine of the critical point is
\[
  z = \frac{ 1 + \rho^2 \, { \cos (\omega)}^2 \pm \sin (\omega) \, \sqrt{ 
   { \sin (\omega) }^2 + {(1 - \rho^2)}^2 \, { \cos (\omega) }^2 } }{ 2 \, \rho \, \cos (\omega) }.
\]
 Hence, a critical point is given by $\arccos (z)$ when $|z| \leq 1$; in this case, we can evaluate $f(\lambda) = { | \Psi (e^{-i \lambda} ) |}^{2n}$
 at this point and compare to the values at $\lambda = 0, \pi$, thereby determining $\lambda^{\star}$ as the minimizer:
\[
 c = { | \Psi (e^{-i \lambda^{\star}} ) |}^{2n}.
\]
The stabilized Butterworth cycle spectral density is then
\[
    f^{\star} (\lambda) =   \frac{ {|  1 - \rho \, \cos (\omega) \, e^{-i \lambda} |}^{2n}  - 
	c \, {| 1 - 2 \, \rho \, \cos (\omega) \, e^{-i \lambda}  + \rho^2 \, e^{-i 2 \lambda } |}^{2n}  }{   {| 1 - 2 \, \rho \, \cos (\omega) \, e^{-i \lambda}  + \rho^2 \, e^{-i 2 \lambda } |}^{2n} }
 \,   \sigma^2,
\]
 which corresponds to an ARMA($2n$, $2n$) process.  The moving average polynomial can be determined by spectral factorization;
 the autocovariances are simply given by those of the original Butterworth cycle process, only with the variance being decreased by $c$.

\subsection{Balanced Cycle}

The Balanced cyclical process is described in  \cite{HarveyTrimbur2003}.   
 This is described through a formulation in state space, but can be expressed
 in ARMA form as follows.     The order $n$ {\em Balanced cyclical process} is defined via
\begin{align*}
  {(1 - 2 \, \rho \, \cos (\omega) B + \rho^2 \, B^2)}^n \, x_t  & =  .5 \, B^{n-1} \,  \left[ { \left( 1 - \rho \, e^{ i \omega } B \right) }^n  +
	 { \left( 1 - \rho \, e^{- i \omega } B \right) }^n \right]  \, \kappa_t  \\
	& + .5 \, i  \, B^{n-1} \,  \left[ { \left( 1 - \rho \, e^{i \omega} B \right) }^n - { \left( 1 - \rho \, e^{-i \omega} B \right) }^n \right]
\, \kappa^*_t,
\end{align*}
 for $\kappa_t \sim \mbox{WN} (0, \sigma^2)$ and $\kappa_t^* \sim \mbox{WN} (0, \sigma^2)$, independent of one another.  
  The moving averages of $\{ \kappa_t \}$ and $\{ \kappa_t^* \}$ are each real polynomials in $B$ of degree $2n-1$; writing them
  in terms of the complex number $e^{i \omega}$ produces compact expressions that facilitate other calculations.  
  The expressions for the autocovariance function given in \cite{Trimbur2006} can be rewritten as
\begin{align*}
  \gamma_h & =  \rho^h \, \cos (h \omega) \,  \left( \sum_{j=0}^{n-1} \binom{h}{j} \, \alpha_{n-j}  \right)  \, {(1 - \rho^2 )}^{1-2n}  \, \sigma^2 \\
\alpha_k & = {(1 - \rho^2 )}^{n-k} \, \sum_{r=0}^{k-1} \binom{k-1}{r} \, \binom{n-1}{r+n-k} \, \rho^{2r}
\end{align*}
 for $h \geq 0$, where we interpret $\alpha_k$ for $k \leq 0$ as zero (and $\binom{h}{j} = 0$ if $j > h$).  A simple expression for 
 the spectral density can be obtained as follows:   
%
%
  the spectral density of the moving average portion is obtained by substituting $e^{- i \lambda}$ for $B$ in both moving
 average polynomials, followed by taking the squared
 magnitude  and summing both terms.   This yields
\begin{align*}
 & .25 \, { \left|   { \left( 1 - \rho \, e^{ i \omega } e^{-i \lambda} \right) }^n  + { \left( 1 - \rho \, e^{ i \omega } e^{-i \lambda} \right) }^n  \right|}^2
 +    .25 \, { \left|   { \left( 1 - \rho \, e^{ i \omega } e^{-i \lambda} \right) }^n  - { \left( 1 - \rho \, e^{ i \omega } e^{-i \lambda} \right) }^n  \right|}^2
  \\
& = .5 \, {( 1 - 2 \, \rho \, \cos (\omega - \lambda) + \rho^2 )}^n + .5 \, {( 1 - 2 \, \rho \, \cos (\omega + \lambda) + \rho^2 )}^n.
\end{align*} 
  Noting that the two summands comprise the two terms in the squared magnitude of the AR(2) polynomial's frequency response function, we see
 that the spectral density is
\[
 f(\lambda)  = .5 \, \left[ {( 1 - 2 \, \rho \, \cos (\omega - \lambda) + \rho^2 )}^{-n} +  {( 1 - 2 \, \rho \, \cos (\omega + \lambda) + \rho^2 )}^{-n}
  \right] \, \sigma^2.
\]
 These results can be used to determine the  ARMA($2n$,$n$) representation of the process: the autocovariance function corresponding
 to the moving average portion can be expressed as
\[
   \upsilon_h = \cos (\omega h) \, \sum_{k=0}^{n-h} \binom{n}{k+h} \binom{n}{k} {(- \rho)}^{2k+h}
\]
 for $0 \leq h \leq n$, by using the binomial formula on ${(1 - \rho \, e^{i \omega } B)}^n$ to determine the autocovariance generating function.
  Also $\upsilon_h = 0$ for $h > n$, so we can compute the moving average polynomial using spectral factorization.

 Another consequence
 of the simple expression for the spectral density is that the stabilized balanced cycle is easy to determine.  Let
 $ g(\lambda) =  {( 1 - 2 \, \rho \, \cos (\omega - \lambda) + \rho^2 )}^{-n} $, which is uni-modal with minimum value over $[0,\pi]$
 occurring at $\lambda = \pi$ so long as $\omega \in (0, \pi/2)$.  (Otherwise, if $\omega \in [\pi/2, \pi)$, there is a minimum value at 
 $\lambda = 0$ -- but this case is of less interest in practice.)  The maximizer of $g$ is $\omega$ (without loss of generality, suppose
 this is positive), and hence
 $g (\pi) < g (\lambda)$ for all $\lambda \in [0, \pi]$ and $g(-\pi) < g(\lambda)$ for all $\lambda \in [-\pi, \omega]$.  It follows
 that $f(\lambda) = .5 \, (g (\lambda) + g(-\lambda)) \, \sigma^2$ satisfies $f(\pi) < f(\lambda)$ for all $\lambda \in [0,\pi]$.
  Hence $c = f(\pi)$, and the stabilized Balanced cycle spectral density is
\[
  f^{\star} (\lambda) =  \frac{  .5 \, { g(\lambda) }^{-1} +  .5 \, { g(-\lambda) }^{-1} - 
	f(\pi) \, {| 1 - 2 \, \rho \, \cos (\omega) \, e^{-i \lambda}  + \rho^2 \, e^{-i 2 \lambda } |}^{2n}  }{ 
  {| 1 - 2 \, \rho \, \cos (\omega) \, e^{-i \lambda}  + \rho^2 \, e^{-i 2 \lambda } |}^{2n} } \, \sigma^2.
\]
 Again, spectral factorization can be used to determine the moving average corresponding to the numerator.
  But the autocovariance function is the same as the original process, but with the variance decremented by $c = f(\pi)$.







\section{Fractional Seasonal Period}
\label{app:fractional-seasonal-period}
 
\subsection{Seasonal Adjustment of Weekly Time Series}

 \cite{ClevelandScott2007}  discuss a technique for adjusting weekly series based upon locally
   weighted regression,  motivated by the varying presence of either 52 or 53 weeks in a year.        Starting with the SABL method of 
        \cite{ClevelandDunnTerpenning1978}, 
  this non-integer period of seasonality has been handled without classical filters, as in the X-11 procedure.
 The linearized X-11 filter for monthly time series is described in
  \cite{LadirayQuenneville2012};  a simplified,
 heuristic discussion is given in Chapter 3 of 
 \cite{McElroyPolitis2020}).   One proceeds in four steps,
   letting $s$ be the seasonal period:
 \begin{enumerate}
 \item Crude Trend:  apply the symmetric filter  $\Upsilon_{\mbox{trend}} (B) = 
 (1+B) (1+ B + \ldots + B^{s-1}) B^{-s/2}/(2s)$ to get an initial
  trend with both seasonality and high-frequency noise suppressed.
  \item De-trend:  subtract the crude trend from the data.  Equivalently, apply $1 - \Upsilon_{\mbox{trend}} (B)$.
  \item Seasonal:  now apply to the de-trended data an order $p$ seasonal moving average of the form $\Upsilon_{\mbox{seas}} (B) 
   =  {(2p+1)}^{-1} \sum_{j=-p}^p  B^{sj}$.
   \item De-seasonalize:  finally, subtract the extracted seasonal from the data.  
  \end{enumerate}
  Altogether, the seasonal adjustment filter is
\[
   \Upsilon_{\mbox{sa}} (B) =    1 -  \Upsilon_{\mbox{seas}} (B)  \,  \left( 1 - \Upsilon_{\mbox{trend}} (B) \right).
 \]
 This simple approach, which is predicated on an additive decomposition of the data in terms of trend, seasonal, and irregular
  components, is effective in many cases, but requires $s$ to be an even integer.   Actually, the technique can be
   adapted to the case that $s$ is an odd integer (e.g., when removing weekly effects from a daily time series), but
   for non-integer $s$ it is unclear how to proceed.  
   
    For weekly time series this generalization is needed,  because
   the number of weeks occurring in a year is not an integer: on average, there are $365.25$ days in a year
   (this includes leap year, but excludes the corrections occurring every 100 years and every 400 years, for simplicity),
   and dividing by $7$ yields $s = 52.1786$.    
      To handle non-integer $s$, we first observe that in step 1 the crude trend filter is constructed from
   \[
   (1+B) (1+ B + \ldots + B^{s-1}) = \prod_{k=1}^{s/2} (1 - 2 \cos (2 \pi k/s) B + B^2)
   \]
   when $s$ is an even integer.
    This ensures annihilation of all sinusoids of frequency $2 \pi k/s$ for $k = 1, \ldots, s/2$  \citep{McElroyPolitis2020}.
    So for a generalization, consider
\begin{equation}
\label{eq:ub.general}
   U_{n,s} =      \prod_{k=1}^{n} (1 - 2 \cos (2 \pi k/s) B + B^2)
\end{equation}
 for integer  $n \leq [s/2]$,  where $[ \cdot ]$ denotes the floor function.
  Then define $  \Upsilon_{\mbox{trend}}(B) = c^{-1} \,  B^{-n}   U_{[s/2],s} (B)$, where
   $c$ is a normalization constant such that $\Upsilon_{\mbox{trend}} (1) = 1$.
  It is apparent that such a filter annihilates the desired harmonics,  and   maintains the level of an input time series.
      To modify the seasonal filter, it is useful to study $ 1 - \Upsilon_{\mbox{seas} } (B)$.   Observe that for even integer $s$,
   \begin{align*}
   1 - \Upsilon_{\mbox{seas} } (B) & = {(2p+1)}^{-1} \, \sum_{j=-p}^p  (1 - B^{sj})  
     = {(2p+1)}^{-1} \, \sum_{j=1}^p  (2 - B^{sj} - B^{-sj} )  \\
   1 - B^{sj} & =  (1-B) (1+ B + \ldots + B^{sj-1}) = (1-B^2) \, \prod_{k=1}^{(sj-2)/2} (1 - 2 \cos (2 \pi k/s) B + B^2),
   \end{align*}
   for $j \geq 1$.     This indicates that for non-integer $s$ we can define the filter as
\[
   \Upsilon_{\mbox{seas}} (B) = 1 - {(2p+1)}^{-1} \sum_{j=1}^p
  \left(  (1-B^2) \, U_{ [(s|j|-2)/2], s} (B)  +  (1 - B^{-2}) \, U_{[(s|j|-2)/2], s} (B^{-1}) \right).
   \]
 When $s$ is large, direct computation of the coefficients of $U_{[(s|j|-2)/2],s} (z)$  can generate numerical distortions.
   For better numerical stability, we use the cepstral method for computing the product of polynomials;
    see pp.232-237 of \cite{McElroyPolitis2020} for background.   The $\ell$th cepstral coefficients
     $\tau_{\ell}$ for $\ell \geq 1$ corresponding to the polynomial $1 - 2 \cos (2 \pi k/s) z + z^2$
      are given by
\[
  \tau_{\ell} =  -2 \cos( 2 \pi k \ell/s) / \ell,
  \]
   and hence the cepstral coefficients of $\prod_{k=1}^{[sj-2]/2} (1 - 2 \cos (2 \pi k/s) z + z^2)$
   are given by summing $\tau_{\ell}$ over $k = 1, \ldots, [sj-2]/2$ for any $j = 1, \ldots, p$.
  It is possible to construct weekly seasonal adjustment filters in this way.
    
\subsection{Weekly SARIMA Models }    
    
    A drawback of constructing AC seasonal adjustment filters, as described above,  is that the user has
     little control over the relative degree of smoothing over trend and seasonal dynamics.   The only parameter
     is $p$, which offers little flexibility.   In contrast,  WK filtering offers a wider class of filters corresponding
     to model parameter MLEs, which are adapted to the empirical features in the data.  A classical decomposition
     posits trend, seasonal, and irregular latent processes
     \citep{McElroyPolitis2020},  where there models
     are either estimated through the structural approach or are algebraically derived from a process model
     via a so-called canonical decomposition 
     \citep{HillmerTiao1982}.
     
    The general form of such latent processes is as follows:  let $\{ S^{(1)}_t \}$ be trend, 
    $\{ S^{(2)}_t \}$ be seasonal, and $\{ S^{(3)}_t \}$ the irregular (a white noise), specified as 
\begin{align*}
   { (1 - B) }^d   S_t^{(1)}  & =   \Psi^{(1)} (B)  \epsilon_t^{(1)} \\
    U_{[s/2],s} (B)  S_t^{(2)} & = \Psi^{(2)} (B)  \epsilon_t^{(2)}.
\end{align*}    
The right hand side specification of $\Psi^{(1)} (z)$ and $\Psi^{(2)} (z)$ can correspond to ARMA or SARMA
 models (for example), and typically $d = 1,2$.
 Then from standard theory \citep{bell1984}  the WK filter for the trend-irregular component has
 frequency response function equal to a bounded function times
  ${| U_{ [s/2],s} (e^{-i \lambda}) |}^2$, indicating that all frequencies of the form $\lambda = 2 \pi k/s$
   for integer k   will be annihilated.    
   
   Although such models can be estimated structurally, there may be an advantage to fitting a SARIMA process
   model and obtaining the latent models by decomposition.  To do so, we need an extension of the SARIMA
   model to non-integer periods.   \cite{LadirayMazzi2018} 
   provide a so-called fractional SARIMA model by
   a Taylor series approximation to the differencing operator $1 - B^s$.   To explain their approach,
   let $s = [s] + \langle s \rangle$, so that $\langle s \rangle $ is the fractional remainder.
   By Taylor series expansion around unity,
$  z^{ r}  \approx  1 +  r ( 1 - z)$
   for $0 \leq r \leq 1$.   Therefore
\begin{align*}
  1 - z^s & =  1 - z^{ [s]}  z^{ \langle s \rangle } \approx 1 - z^{[s]}  (1 + \langle s \rangle (1-z))  \\
   &   =  1 - (1 + \langle s \rangle ) z^{ [s]}  + \langle s \rangle  z^{ [s] + 1}.
\end{align*}
   This operator does not annihilate dynamics of frequency $\lambda = 2 \pi k/s$,  except in the basic case
    that $\langle s \rangle = 0$, so  this does not seem to be the correct approach.   The deficiency can be
    rectified by taking infinitely many terms in the Taylor series expansion, but this is of no practical use
    because the differencing operator will no longer be a polynomial.   In contrast, the approach adopted
    above in constructing $U_{n,s} (z)$ is to begin with the frequencies one seeks to annihilate, and then
    incorporate all the minimal degree polynomials that accomplish this nullification.  
    
    We describe an alternative approach to constructing fractional period SARIMA models, which we hope to implement in future versions of \pkg{Ecce Signum}.
    Recall that
\[
  (1- z) U_{[s/2],s} (z) = 1 - z^s
 \]
  when $s$ is an even integer.   Plugging in $z = \vartheta^{1/s} B$ for some $\vartheta \in (0,1]$, we see that
   operators of the form $1 - \vartheta B^s$ appearing in SARIMA model specifications can be replaced,
   for fractional $s$, by
\begin{equation}
\label{eq:fractional-operator}
    (1 - \vartheta^{1/s} B)  U_{[s/2],s}  (  \vartheta^{1/s} B).
\end{equation}
  For instance, a SARIMA(p,q,1,1) could be written as
  \[
   {(1 - B) }^2  U_{[s/2],s} (B)  \phi (B)   (1 - \Phi^{1/s} B) U_{[s/2],s} ( \Phi^{1/s} B)    X_t =  
    \theta(B) (1 - \Theta^{1/s} B) U_{[s/2],s} ( \Theta^{1/s} B)  \epsilon_t.
\]
    This does not work if $\Phi$ or $\Theta$ take negative values.   Also, for higher order
    specifications,  we need to include more and more terms of type (\ref{eq:fractional-operator}); 
    however, these would be used only rarely.






\section{Functionality}
\label{app:function}

There are over 70 functions in the \pkg{Ecce Signum} package,  which fall into three categories: 
 (i) general calculation functions, (ii) algorithms for multivariate time
 series, and (iii) functions particular to \pkg{sigex} model structures.   In the first category are functions for performing algebraic 
operations, as well as some generic time series functionality; these have no prefix in their nomenclature.  In the second category
 are functions written for multivariate time series to compute likelihoods or generate forecasts, and do not presume any particular
 model forms but instead are based on the autocovariance function.   These have an ``mvar" prefix in their nomenclature.
 The third category works with the presumed model and parameter structures of \pkg{sigex}, and have a ``sigex" prefix in their
 nomenclature.   Tables \ref{tab:function1},  \ref{tab:function2}, \ref{tab:function3},   \ref{tab:function4},   \ref{tab:function5},
  and \ref{tab:function6}
 provide an alphabetical list  (prefix is ignored in the alphabetization) of the package's functions, along with a brief description
as well as which functions it calls and is called by.

\newpage

\begin{table}[htb!]
\centering
\begin{small}
\begin{tabular}{llll}
 \hline Function & Purpose &  Calls  &  Called By \\
\hline
\code{sigex.acf}  &   Compute the autocovariance function &  \code{ARMAauto}	&    \code{sigex.cast}  \\
	& \quad of a differenced latent component  &  \code{sigex.canonize} &  \code{sigex.lik} 	\\
	&	& \code{sigex.getcycle} & \code{sigex.midcast}	\\
	&	&	\code{polymulMat} & 	\\
	&	& \code{polymult} &  \code{sigex.resid}	\\
	&	& \code{polysum} &   \code{sigex.signal}	\\
	&	& \code{specFact} &   \code{sigex.whittle}  \\
	&	& \code{specFactmvar} &  \\
	&	& \code{VARMAauto} &   \\
\code{sigex.add} &  Adds a  latent component to model  &   & \code{sigex.reduce}  \\
\code{sigex.adhocextract} &   Compute signal extractions and MSE  & \code{sigex.cast}  &      \\
	& \quad from an ad hoc filter  & \code{sigex.midcast} &  \\
	&		&  \code{sigex.psi2par}  &  \\
\code{ARMAauto} & Compute   autocovariances of   ARMA  &   &  \code{sigex.acf} \\
	&	&	& \code{sigex.momfit} \\
\code{sigex.blocktoep} &  Generates lower triangular array  &   &  \code{sigex.signal}  \\
	& \quad  for block Toeplitz matrix &  &  \\
\code{sigex.bundle} &  Writes fitted model information to a list &  &  \\
\code{sigex.canonize} &   Computes canonization of  &  \code{polymult}  &   \code{sigex.acf}	\\
	&  \quad a given ARMA process 	&  \code{polysum}   &  \code{sigex.momfit}	\\
	&	& \code{sigex.specFact} &  \code{sigex.spectra} 	\\
\code{sigex.cast} &   Computes forecasts and aftcasts  &  \code{sigex.acf} &  \code{sigex.wkextract}  \\
	& \quad  without uncertainty  &  \code{sigex.delta} & 	\\
	&	& \code{mvar.forecast} & 	\\
	&	& \code{sigex.param2gcd}	&	\\
	&	& \code{sigex.zeta2par}	&	\\
	&	& \code{sigex.zetalen}  &	\\
\code{sigex.castextract} &  Formats casts for plotting &    & 		\\	
\code{sigex.conditions} &  Computes condition number & \code{getGCD}    &  \code{sigex.reduce}  \\
		&  \quad  for a covariance matrix  	&   \code{sigex.param2gcd}  &  	\\
\code{sigex.constrainreg} &   Determines constraint matrix for regressors  &  \code{sigex.default} & 	\\
	&	& \code{sigex.par2psi}	&	\\
\code{daily2monthly} &  Converts a daily time series   &  \code{date2day} &  	\\
	&  \quad to a monthly time series		& \code{day2date}  & 	\\
	&		& \code{day2week} & 	\\
\code{sigex.daily2weekly} & 	Embeds a daily time series  &  \code{date2day} &  \\
		& \quad as a weekly time series 	& \code{day2week} & 	\\
\hline
\end{tabular}
\caption{\baselineskip=10pt   \pkg{Ecce Signum} Function Taxonomy. }
\label{tab:function1}
\end{small}
\end{table}

\begin{table}[htb!]
\centering
\begin{small}
\begin{tabular}{llll}
 \hline Function & Purpose &  Calls  &  Called By \\
\hline
\code{date2day} &  Converts month-day-year date  & 		&	\code{day2date} \\
	& \quad to a  day index &		&	\code{day2week} 	\\
	&		&	&	\code{sigex.daily2weekly} \\
\code{datefinder} &  Finds a holiday date &   \code{date2day}  &  \\
	&  \quad  in month-day-year format	&   \code{day2date}  &	\\
	&	&    \code{day2week} &	\\
\code{day2date} &   Converts a day index to a date  &  \code{date2day} &  	\\
	&   \quad  in month-day-year format  & 		&	\\
\code{day2week} &    Converts a date in month-day-year  	& 	\code{date2day} &  \code{sigex.daily2weekly} 	\\
	&	\quad  format and finds the day of week  & 	&	\\
\code{sigex.default} &  Initializes param with zeroes  &  \code{sigex.eta2psi}  &   \code{sigex.mvar2uvar} \\
	&	&  \code{sigex.psi2par}  &  \\
	&	&  \code{sigex.zetalen} & 	\\
\code{sigex.delta} &  Computes differencing polynomial &  \code{polymult} & \code{sigex.cast}  \\
	& \quad  with factors omitted &  	& \code{sigex.frf}  \\
	&	&	& \code{sigex.lik} \\
	&	&	& \code{sigex.lpmse}  \\
	&	&	& \code{sigex.meaninit} \\
	&	&	& \code{sigex.midcast} \\
	&	&	& \code{sigex.momfit} \\
	&	&	& \code{sigex.reg} \\
	&	&	& \code{sigex.resid} \\
	&	&	& \code{sigex.signal} \\
	&	&	& \code{sigex.whittle} \\
	&	&	& \code{sigex.wkmse} \\
\code{sigex.eta2psi} &  Transforms  $\eta$ to $\psi$  &     &  \code{sigex.default}  \\
	&	&	&  \code{sigex.mlefit}  \\
	&	&	& \code{sigex.tstats} \\
\code{sigex.extract} &  Computes signal extraction estimates  &   &   \\
	& \quad   with two standard errors &  &  \\
\code{sigex.fixed} & Computes all non-trend fixed regression effects   &   &  \\
\code{mvar.forecast} &   Computes   forecasts and predictors  &   &  \code{sigex.cast} \\
\code{sigex.frf} &  Computes signal extraction filter &  \code{sigex.delta}  &  \code{sigex.wk}   \\
	& \quad frequency response function  & \code{sigex.spectra} &	\\
\code{sigex.gausscheck} &  Tests for normality &   &  \\
\code{sigex.getcycle} &  Computes AR and MA polynomials  &  \code{polymult} & \code{sigex.acf}  \\
	& \quad for Butterworth and Balanced cycles &   	&  \code{sigex.momfit} \\
	&	&	& \code{sigex.spectra} \\
 \hline
\end{tabular}
\caption{\baselineskip=10pt   \pkg{Ecce Signum} Function Taxonomy. }
\label{tab:function2}
\end{small}
\end{table}

\begin{table}[htb!]
\centering
\begin{small}
\begin{tabular}{llll}
 \hline Function & Purpose &  Calls  &  Called By \\
\hline
\code{sigex.getfrf} &  Computes and plots signal   &  \code{sigex.frf} & \\
	& \quad frequency response function & 	&	\\
\code{getGCD} &  Computes the generalized   &     &  \code{sigex.conditions}   \\
	&	\quad  Cholesky decomposition &	&   	\code{sigex.momfit}    \\
\code{gethol} &  Generates holiday regressors from holiday dates &   \code{date2day} &   \\
\code{sigex.glr}  &   Computes log likelihood ratio statistic &  \code{sigex.lik} &  \\
	& \quad for nested models   &   &  \\
\code{sigex.graph} &   Graphs  signal extraction estimates  &   &  \\
	&  \quad to an existing time series plot &   &  \\
\code{sigex.hi2low} &  Embeds a  high frequency filter  &   &  \\
  &   \quad  as a low frequency filter   &   & \\
\code{sigex.i2rag} &  Generate list of missing value indices  &    & \code{sigex.midcast} 	\\
\code{sigex.lik} &   Computes  Gaussian divergence of model  &  \code{sigex.acf}  &  \code{sigex.glr}  	\\
	&	& \code{sigex.delta} 	&	\code{sigex.mlefit}   \\
	&	& \code{mvar.midcast}	&	\\
	&	& \code{sigex.param2gcd} &	\\
	&	& \code{sigex.zeta2par}	&	\\
	&	& \code{sigex.zetalen}	&	\\  \code{sigex.load}  &  Loads data into a time series object &   &   \\
\code{sigex.lpfiltering} &   Computes trend and cycle extractions  &  \code{sigex.lpmse} &  \\
	& \quad by combining with a low-pass filter  &  \code{sigex.lpwk}  &   \\
\code{sigex.lpmse} &   Computes trend and cycle MSE  &   \code{sigex.delta}	&  \code{sigex.lpfiltering}   \\
	& \quad by combining with a low-pass filter  &  \code{sigex.spectra}  &   \\
 	&	&  \code{sigex.wkmse}  \\
\code{sigex.lpwk} &    Computes trend and cycle filters  &  \code{sigex.wk} &   \code{sigex.lpfiltering}   \\
	& \quad by combining with a low-pass filter  &  &   \\
\code{sigex.meaninit} &   Adds trend regressors to an existing model &  \code{sigex.delta} & \code{sigex.reduce} 	\\
	&	&	\code{sigex.whichtrend}  & 	\\
\code{mvar.midcast} &   Computes  multi-step imputations and  &  	& \code{sigex.lik}  \\
	& \quad predictors of a multivariate process  &		&  \code{sigex.midcast}  \\
	&  \quad via Levinson-Durbin algorithm 		&	& \code{sigex.resid}  \\
	&  \quad with  missing values			&	&    \\
\hline
\end{tabular}
\caption{\baselineskip=10pt   \pkg{Ecce Signum} Function Taxonomy. }
\label{tab:function3}
\end{small}
\end{table}

\begin{table}[htb!]
\centering
\begin{small}
\begin{tabular}{llll}
 \hline Function & Purpose &  Calls  &  Called By \\
\hline
\code{sigex.midcast} &   Computes predictors for variables   &   \code{sigex.acf}  &  \code{sigex.wkextract}  \\
	&  \quad  at various indices  &   \code{sigex.delta}  &  \\
	&	& \code{sigex.i2rag}	&	\\
	&	& \code{mvar.midcast}	& 	\\
	&	&  \code{sigex.param2gcd}  &		\\
	&	& \code{sigex.zeta2par} 	&	\\
	&	& \code{sigex.zetalen}	&	\\
\code{sigex.mlefit} &   Fits  model to the data via MLE &  \code{sigex.eta2psi}    &   \\
	&	&    \code{sigex.lik}   &  \\
	&	&   \code{sigex.par2psi}    &  \\
	&	&  \code{sigex.psi2eta}  &  \\
	&	&  \code{sigex.psi2par}  &  \\
\code{sigex.momfit} &  Computes initial parameter estimates  &  \code{ARMAauto} & 		\\
	& \quad   by method of moments  &  \code{sigex.canonize}   & 		\\
	&	&  \code{sigex.delta}  &  \\
	&	&  \code{sigex.getcycle} &  \\
	& 	& \code{getGCD} & 	\\
	&	&  \code{polymult} & \\
	&	&  \code{polysum}  &   \\
\code{sigex.mvar2uvar} &  Transform multivariate parameters 	 	&  \code{sigex.default}  &  \code{sigex.precision}  \\
	& \quad to implied univariate form  & 	&	\\
\code{var2.par2pre} &   Computes VAR pre-parameters   &  \code{VARMAauto} & \code{sigex.par2zeta}  \\
\code{sigex.par2psi} &   Transforms param to $\psi$  &  \code{sigex.par2zeta} &  \code{sigex.mlefit}  \\	
	&	&	&  \code{sigex.reduce}  \\
\code{sigex.par2zeta} &  Transforms param to $\zeta$  &  \code{var2.par2pre}   & \code{sigex.par2psi}  \\
\code{sigex.param2gcd} &  Maps real vector to  a &   &  \code{sigex.cast} \\
	& \quad unit lower triangular matrix & 	&   \code{sigex.conditions} \\
 	&	&	& \code{sigex.lik}  \\
	&	&	& \code{sigex.midcast} \\
	&	&	& \code{sigex.psi2par}  \\
	&	&	&  \code{sigex.resid}  \\
	&	&	& \code{sigex.whittle}  \\
	&	&	& \code{sigex.zeta2par}  \\
\code{polymulMat} &  Computes the product of  & 	&	\code{sigex.acf} \\
	&	\quad two matrix polynomials &	&	\\
\code{polymult} &  Computes the product of two polynomials &	 & \code{sigex.acf}  \\
	&	&	& \code{sigex.canonize} \\
	&	&	& \code{sigex.delta} \\
	&	&	& \code{sigex.getcycle} \\
	&	&	& \code{sigex.momfit} \\
	&	&	& \code{sigex.spectra} \\
\code{polysum} &  Compute the sum of two polynomials  &  	&  \code{sigex.acf}  \\
	&	&	& \code{sigex.canonize} \\
	&	&	& \code{sigex.momfit} \\
	&	&	& \code{sigex.spectra} \\
\code{sigex.portmanteau}  &   Portmanteau  tests for autocorrelation &   &  \\
\code{var2.pre2par} &   Generates a stable  VAR process  &  	& \code{sigex.zeta2par} \\
\hline
\end{tabular}
\caption{\baselineskip=10pt   \pkg{Ecce Signum} Function Taxonomy. }
\label{tab:function4}
\end{small}
\end{table}

\begin{table}[htb!]
\centering
\begin{small}
\begin{tabular}{llll}
 \hline Function & Purpose &  Calls  &  Called By \\
\hline
\code{sigex.precision} &   Compares signal extraction MSE arising from &  \code{sigex.mvar2uvar}   &   \\
	& \quad 	multivariate fit with the implied univariate fit  &  \code{sigex.signal}  &  \\
\code{sigex.prep}  &  Applies   preliminary transformations to the data  &  \code{sigex.transform}  &  \\
\code{sigex.psi2eta} &  Transforms $\psi$ to $\eta$  &   &   \code{sigex.mlefit}	\\
\code{sigex.psi2par} & Transforms $\psi$ to param &  \code{sigex.param2gcd}	&    \code{sigex.default}  \\
	&	& \code{sigex.zeta2par}  &  \code{sigex.mlefit}  \\
	&	& \code{sigex.zetalen} &  \code{sigex.wkextract}  \\
\code{sigex.reduce} &   Obtain a reduced rank model  &   \code{sigex.add}  &   \\
	&	&  \code{sigex.conditions}  &  \\
	&	&  \code{sigex.meaninit} &  \\
	&	& \code{sigex.par2psi} &   \\
	& 	&   \code{sigex.renderpd}  &   \\
\code{sigex.reg} &   Adds regressors to an existing model  &  \code{sigex.delta}  &   \\
\code{sigex.renderpd} &   Modifies a covariance matrix so that it is pd &  	& \code{sigex.reduce}  \\
\code{sigex.resid} &   Computes residuals from model's   &  \code{sigex.acf} &  \\
	&  \quad  Gaussian likelihood 	&	\code{sigex.delta}  & 		\\
	&	&	\code{mvar.midcast} & 		\\
	&	&	 \code{sigex.param2gcd} & 	\\
	&	&	 \code{sigex.zeta2par} &	\\
	&	&	 \code{sigex.zetalen} &		\\
 \code{sigex.signal} &   Computes signal extraction matrix  &   \code{sigex.acf} &   \code{sigex.precision} \\
	& \quad  and error covariance matrix  &   \code{sigex.blocktoep} &  \\
	&	&	 \code{sigex.delta} &   \\
\code{sigex.signalcheck} &   Computes model-based  &  \code{sigex.acf} & 	\\
	&	\quad signal extraction diagnostics &  \code{sigex.delta} &	\\
\code{sigex.sim} &  Simulates  a stochastic process &   \code{sigex.acf} & 	\\
	&	& \code{sigex.delta} &  	\\
	&	& \code{sigex.param2gcd} 	&  \\
	&	& \code{sigex.zeta2par}	& \\
	&	& \code{sigex.zetalen} & 	\\
 \code{sigex.specar} &  Plot AR spectrum of a time series  &   &    \\
\code{specFact} &  Compute the spectral factorization of a  &  	&  \code{sigex.acf}  \\
	&   \quad given univariate pd sequence	&	& \code{sigex.canonize} \\
	&	&	& \code{sigex.momfit} \\
	&	&	& \code{sigex.spectra} \\
\code{specFactmvar} & Compute the spectral factorization of a &  & \code{sigex.acf}  \\
	&	 \quad given multivariate pd sequence &  & \code{sigex.spectra} \\
\code{sigex.spectra}  &    Computes scalar part of spectrum  &  \code{sigex.canonize}  &    \code{sigex.frf}  \\
	& \quad of a differenced latent component  & \code{sigex.getcycle}  &  \code{sigex.lpmse} 	\\
	&	&  \code{polymult}  & \code{sigex.whittle}	\\
	&	&   \code{polysum} &  \code{sigex.wkmse}	\\
	&	& \code{specFact} &      \\
	&	& \code{specFactmvar} &  \\
\hline
\end{tabular}
\caption{\baselineskip=10pt   \pkg{Ecce Signum} Function Taxonomy. }
\label{tab:function5}
\end{small}
\end{table}

\begin{table}[htb!]
\centering
\begin{small}
\begin{tabular}{llll}
 \hline Function & Purpose &  Calls  &  Called By \\
\code{sigex.transform} &  Applies aggregation and  transformations  &  & \code{sigex.prep}  \\
 \code{sigex.tstats} &  Computes t statistics for parameter estimates &  \code{sigex.eta2psi}  &  \\
 \code{ubgenerator} &  Computes the product of unit root	 &   	& \code{x11filters} \\
 	&	\quad  differencing operators	&	&	\\
\code{VARMAauto} & Computes autocovariances of VARMA  &   &  \code{sigex.acf} \\
	&	&	&  \code{var2.par2pre} \\
\code{sigex.weekly2daily} &  De-embeds a weekly time series &  \code{date2day} 	&	\\
	& \quad as a daily time series &   \code{day2week}	&	\\
	&	& \code{sigex.load}	&	\\
\code{weekly2date} &  Obtains start and end dates for  &   \code{date2day}  &      \\
	&	\quad a weekly time series 	& \code{day2date}   &		\\
	&						&  \code{day2week}  &		\\
\code{sigex.whichtrend} &   Determines which component is for trend  &  & \code{sigex.meaninit} \\
\code{sigex.whittle} &   Computes Whittle likelihood of model   &  \code{sigex.acf}  &  \code{sigex.mlefit}  	\\
	&	& \code{sigex.delta} 	&	\\
	&	& \code{sigex.param2gcd} &	\\
	&	& \code{sigex.spectra}	&	\\
	&	& \code{sigex.zeta2par}	&	\\
	&	& \code{sigex.zetalen}	&	\\
\code{sigex.wk}  &    Computes signal extraction filter &  \code{sigex.frf}	& \code{sigex.lpwk} \\
	&	\quad  coefficients and MSE  &  \code{sigex.wkmse} 	&  \code{sigex.wkextract} \\
\code{sigex.wkextract} &   Computes signal extractions   &  \code{sigex.cast} &    \\
	& \quad and MSE via WK method  & \code{sigex.midcast} &  \\
	&	& \code{sigex.psi2par}	&	\\
	&	& \code{sigex.wk} &	\\
\code{sigex.wkmse} &   Computes signal extraction error &    \code{sigex.delta}   &  \code{sigex.lpmse}   \\
	& \quad  spectrum from bi-infinite sample  &   \code{sigex.spectra}  &  \code{sigex.wk}   \\
  \code{x11filters} &  Generates x11 trend, seasonal, and	&  \code{polymult}  &	\\
	&	 \quad  seasonal adjustment filters	&  \code{ubgenerator}	&	\\	
\code{sigex.zeta2par} &  Transform  $\zeta$ to param  &  \code{var2.pre2par}  &   \code{sigex.cast}  \\
	&	&	\code{sigex.param2gcd} &   \code{sigex.lik}  \\
	&	&	& \code{sigex.midcast} \\
	&	&	&  \code{sigex.psi2par} \\
	&	&	& \code{sigex.resid} \\
	&	&	& \code{sigex.whittle}  \\
\code{sigex.zetalen} &   Computes the length of  $\zeta$ &  	& \code{sigex.cast} \\
	&	&	& \code{sigex.default}  \\
	&	&	& \code{sigex.lik}  \\
	&	&	& \code{sigex.midcast} \\
	&	&	& \code{sigex.psi2par}  \\
	&	&	&  \code{sigex.resid}  \\
	&	&	& \code{sigex.whittle}  \\
	\hline
\end{tabular}
\caption{\baselineskip=10pt   \pkg{Ecce Signum} Function Taxonomy. }
\label{tab:function6}
\end{small}
\end{table}

\end{document}